\documentclass[lettersize,journal]{IEEEtran}

\usepackage{cite}
\usepackage{amsmath,amssymb,amsfonts}
\usepackage{graphicx}
\usepackage{textcomp}
\usepackage{stfloats}
\usepackage{xcolor}
\usepackage{array}
\usepackage{url}
\usepackage{verbatim}
\usepackage{balance}
\def\BibTeX{{\rm B\kern-.05em{\sc i\kern-.025em b}\kern-.08em
		T\kern-.1667em\lower.7ex\hbox{E}\kern-.125emX}}

\usepackage{subfigure}
\usepackage{epstopdf}

\usepackage{algorithmic}
\usepackage[ruled,vlined]{algorithm2e}

\begin{document}

\title{Environment-Aware IRS Deployment via Channel Knowledge Map: Joint  Sensing-Communications Coverage Optimization}
	
\author{Yilong Chen, \IEEEmembership{Graduate Student Member, IEEE}, Zixiang Ren, \IEEEmembership{Graduate Student Member, IEEE}, \\ Jie Xu, \IEEEmembership{Fellow, IEEE}, and Rui Zhang, \IEEEmembership{Fellow, IEEE} \vspace{-0.4cm}
	\thanks{Y. Chen, Z. Ren, and J. Xu are with School of Science and Engineering, Shenzhen Future Network of Intelligence Institute (FNii-Shenzhen), and Guangdong Provincial Key Laboratory of Future Networks of Intelligence, The Chinese University of Hong Kong, Shenzhen, Guangdong 518172, China (e-mail: yilongchen@link.cuhk.edu.cn; rzx66@mail.ustc.edu.cn; xujie@cuhk.edu.cn).}
	\thanks{R. Zhang is with the Department of Electrical and Computer Engineering, National University of Singapore, Singapore 117583 (e-mail: elezhang@nus.edu.sg).}
	\thanks{J. Xu is the corresponding author.}
}

\maketitle

\begin{abstract}
	This paper studies the intelligent reflecting surface (IRS) deployment optimization problem for IRS-enabled integrated sensing and communications (ISAC) systems, in which multiple IRSs are strategically deployed at candidate locations to assist a base station (BS) to enhance the coverage of both sensing and communications. We present an environment-aware IRS deployment design via exploiting the channel knowledge map (CKM), which provides the channel state information (CSI) between each candidate IRS location and BS or targeted sensing/communication points. Based on the obtained CSI from CKM, we optimize the deployment of IRSs, jointly with the BS's transmit beamforming and IRSs' reflective beamforming during operation, with the objective of minimizing the system cost, while guaranteeing the minimum illumination power requirements at sensing areas and the minimum signal-to-noise ratio (SNR) requirements at communication areas. In particular, we consider two cases when the IRSs’ reflective beamforming optimization can be implemented dynamically in real time and quasi-stationarily over the whole operation period, respectively. For both cases, the joint IRS deployment and transmit/reflective beamforming designs are formulated as mixed-integer non-convex optimization problems, which are solved via the successive convex approximation (SCA)-based relax-and-bound method. Specifically, we first relax the binary IRS deployment indicators into continuous variables, then find converged solutions via SCA, and finally round relaxed indicators back to binary values. Numerical results demonstrate the effectiveness of our proposed algorithms in reducing the system cost while meeting the sensing and communication requirements.
\end{abstract}

\begin{IEEEkeywords}
	Intelligent reflecting surface (IRS), integrated sensing and communications (ISAC), channel knowledge map (CKM), IRS deployment, coverage enhancement.
\end{IEEEkeywords}

\section{Introduction}

Integrated sensing and communications (ISAC) has been recognized an important usage scenario for future sixth-generation (6G) wireless networks \cite{tong20226g}, which leverages wireless signals for the dual purpose of information delivery and environment/target sensing \cite{cui2021integrating, liu2022Integrated, hua2023optimal}, enabling diverse intelligent applications, such as smart homes, smart logistics, industrial automation, and smart healthcare \cite{tong20226g}. To realize this vision, it is crucial to ensure seamless sensing and communication coverage across potential ISAC deployment scenarios.
However, this objective presents significant challenges due to the densely distributed obstacles and scatterers in practical wireless environments. These physical obstructions may block line-of-sight (LoS) propagation, cause severe deep fading, and ultimately degrade both sensing and communication performances.

To address these challenges, the intelligent reflecting surface (IRS) technology has gained considerable attention \cite{wu2021intelligent, mei2022intelligent, wu2024intelligent}. 
An IRS consists of an array of tunable passive elements that can dynamically manipulate the magnitude and phase of incident signals. This enables real-time reconfiguration of the wireless propagation environment, thereby improving both sensing and communication performance in ISAC systems.
Different from  conventional base stations (BSs) and relays, IRS eliminates the need for dedicated radio frequency (RF) chains, offering substantial advantages in terms of hardware cost reduction and energy efficiency enhancement. The low-cost nature of IRS renders it highly suitable for large-scale deployments in future ISAC scenarios.

To effectively harness IRSs for ISAC coverage enhancement, it is essential to carefully design IRS deployment based on wireless propagation environment characteristics.
In the literature, extensive research efforts have been devoted to the deployment design of single/multiple IRSs for wireless communication systems, where simplified LoS or stochastic channel models are considered \cite{you2021wireless, mu2021joint, lu2021aerial, bai2022robust, cheng2022ris, kang2022IRS, fu2023active, kang2023double, feng2023joint, ling2021placement, huang2022placement, efrem2023joint, zhang2024irs}.
For instance, the authors in \cite{you2021wireless, mu2021joint, lu2021aerial, bai2022robust, cheng2022ris, kang2022IRS} investigated the deployment design of a single IRS to enhance the communication system performance, by considering the scenarios with a single communication user (CU) \cite{you2021wireless, bai2022robust, cheng2022ris, kang2022IRS} and multiple CUs \cite{mu2021joint,lu2021aerial}, respectively. In particular, \cite{lu2021aerial} jointly optimized the IRS deployment together with the BS transmit beamforming and IRS reflective beamforming to maximize the minimum communication signal-to-noise ratio (SNR) across a targeted area. \cite{bai2022robust} designed the IRS deployment to enhance the achievable secrecy rate of a legitimate CU in the presence of an eavesdropper. \cite{cheng2022ris} further exploited the IRS plane rotation as an additional design degree of freedom to enhance the communication performance.
Subsequent research extended the IRS deployment strategies to double-IRS cooperative systems, in which the dual-reflecting surfaces either jointly enable cascaded reflections for extending coverage to distant areas \cite{fu2023active,kang2023double}, or operate at separate locations to serve distinct coverage regions \cite{feng2023joint}.
For both scenarios, the deployment of double IRSs was optimized to maximize the communication rate. 
In addition, more recent research further investigated the deployment of multiple IRSs to enhance the communication performance of multiple CUs \cite{ling2021placement, huang2022placement, efrem2023joint, zhang2024irs}. For instance, \cite{huang2022placement} minimized the number of deployed IRSs (and thus the system cost) while ensuring the average achievable data rate at each CU, \cite{efrem2023joint} minimized the outage probability of a full-duplex communication system subject to constraints on the maximum number of deployed IRSs and reflecting elements, and \cite{zhang2024irs} demonstrated that allocating reflecting elements across multiple distributed IRSs yields increased communication SNR as compared to deploying a co-located IRS with the same total number of reflecting elements.

Despite these progresses, the above works on IRS deployment have the following critical limitations for practical ISAC implementation. First, these works predominantly focused on wireless communication systems by adopting simplified LoS or stochastic channel models, which fail to capture the environment-specific propagation characteristics of practical sensing and communication scenarios.
More importantly, while wireless communication benefits from combined multi-path channels (including both LoS and non-line-of-sight (NLoS) links), IRS-enabled sensing primarily depends on establishing virtual LoS links toward sensing targets \cite{song2022joint, song2024overview}. This fundamental difference renders conventional communication-oriented IRS deployment strategies unsuitable for ISAC applications.
Furthermore, prior studies \cite{ling2021placement, huang2022placement, efrem2023joint, zhang2024irs} primarily considered systems with single-antenna BS configurations, where the IRS reflective beamforming design can be simplified by setting each reflecting element's phase shift to align with its corresponding channel coefficient. This becomes inadequate for the scenarios with multi-antenna BS, where the system performance critically depends on the joint optimization of BS transmit beamforming and IRS reflective beamforming. The coupling between these design parameters introduces significant complexity, making the IRS deployment problem more challenging.

Different from prior works focusing solely on communication enhancement, this paper investigates the environment-aware IRS deployment optimization for ISAC systems.
Specifically, multiple IRSs are strategically deployed at predefined candidate locations, in order to cooperatively enhance the sensing and communication coverage of a BS for areas of interest. 
In particular, instead of considering LoS or stochastic channel models as in prior works, we leverage the channel knowledge map (CKM) to obtain environment-specific channel state information (CSI) to enable environment-aware IRS deployment design. In general, CKM is a database tagged with locations of transceivers, which provides location-specific channel knowledge useful to enhance environment-awareness and facilitate the acquisition of location-specific CSI \cite{zeng2021toward}. 
In practice, CKM can be constructed via various techniques like ray-tracing, interpolation, and wireless radiance fields \cite{zeng2024tutorial}. 
CKM enables the CSI prediction at unexplored locations, particularly in complex environments like urban and indoor scenarios, where exhaustive CSI measurements are impractical.
In this paper, we focus on the utilization of CKM for environment-aware IRS deployment. Therefore, it is assumed that with CKM, we obtain location-specific channel knowledge, including multi-path channel gains, delays, and angles of arrival/departure (AoAs/AoDs) between varying transmitter and receiver locations, which provides the point-to-point CSI between each candidate IRS location to the BS or sensing/communication areas.\footnote{How to accurately construct such CKM based on prior channel measurements and physical environment information is an important task that is out of the scope of this paper. In addition, how to analyze the effect of imperfect CKM on the performance of IRS deployment is another interesting problem that is left for future work.} Under this setup, we aim to optimize the deployment of multiple IRSs over a set of candidate locations, jointly with the transmit beamforming at the BS and the reflective beamforming at the IRSs during operation, to minimize the system cost while ensuring both sensing and communication coverage. 
The main results of this paper are summarized as follows. 

\begin{itemize}
	
\item Our objective is to minimize the weighted sum of the number of deployed IRSs and the system power consumption during operation, while ensuring the sensing illumination power and communication SNR requirements across the sensing and communication areas, respectively.
This is a mixed-time-scale problem that is difficult to solve, as the IRS deployment and transmit/reflective beamforming are implemented at different time scales in general. 

\item As the analog reflective beamforming at IRSs may operate at a relatively longer time scale than that for the digital transmit beamforming at BS, we consider two cases with dynamic and quasi-stationary IRS operations, respectively. For the former case, the IRSs' reflective beamforming is adjustable in real time together with the BS's transmit beamforming, while for the latter case, the reflective beamforming is only tunable at the beginning of the whole operation period. 
For both cases, we use a series of binary indicators to represent the IRS deployment over a set of candidate deployment locations. Accordingly, we formulate the joint IRS deployment and transmit/reflective beamforming optimization as mixed-integer non-linear optimization problems that are challenging to solve due to their non-convexity and the coupling of IRS deployment indicators and beamforming vectors. 

\item We propose successive convex approximation (SCA)-based relax-and-round algorithms to efficiently solve the formulated problems. Specifically, we first relax the binary IRS deployment indicators into continuous IRS deployment weights, which are jointly optimized together with the IRS reflective beamforming and BS transmit beamforming by utilizing the SCA technique. After obtaining the relaxed IRS deployment weight, we develop a greedy search-based rounding process to recover the binary IRS deployment indicators, and then determine the corresponding IRS reflective beamforming and BS transmit beamforming.

\item To facilitate practical implementation, we further present a heuristic design with channel-based deployment (CBD) to reduce the computation complexity. 
The CBD-based algorithm determines the weights for IRS deployment based on the channel condition between each candidate IRS location to the sensing and communication areas. Then, based on the IRS deployment weight, a rounding process is employed to recover the binary IRS deployment indicators. 

\item Finally, numerical results are conducted to validate the effectiveness of our proposed IRS deployment algorithms in reducing the system cost while ensuring sensing and communication coverage. It is shown that our proposed SCA-based design achieves adaptive IRS deployment at candidate sites to meet various sensing and communication coverage requirements, via properly balancing between minimizing the capital and operational costs. The heuristic design with CBD is shown to achieve a comparable performance to the SCA-based algorithm with lower time complexity. Both algorithms are shown to outperform the benchmark IRS deployment design with random reflective beamforming (RRB).
\end{itemize}


Notice that there is a prior work \cite{fu2024multi} studying the deployment of multiple IRSs for communication coverage enhancement by leveraging large-scale channel knowledge. The authors in \cite{fu2024multi} proposed an efficient method to approximate the average channel power gain from the BS to targeted location grids through candidate IRSs, and accordingly optimized the IRS deployment cost via omitting the reflective beamforming design at IRSs by approximating the reflection power gain to be proportional to the square of the number of IRS reflecting elements.
By contrast, this paper considers both sensing and communication coverage for IRS-enabled ISAC systems, and explicitly considers the IRSs' reflective beamforming design in the joint optimization problem by exploiting CKM to obtain point-to-point CSI among varying locations.

The remainder of this paper is organized as follows. Section II presents the IRS-enabled ISAC system model, and then formulates the joint IRS deployment and transmit/reflective beamforming optimization problems for two cases with quasi-stationary and dynamic IRS reflective beamforming, respectively. Sections III and IV propose SCA-based relax-and-round algorithms for the formulated problems in the two cases, respectively. Section V proposes the CBD heuristic design and the RRB benchmark.
Section VI provides numerical results to validate the performance of our proposed designs. Finally, Section VII concludes this paper.

\textit{Notations:} Boldface letters are used for vectors (lower-case) and matrices (upper-case). For a square matrix \(\mathbf{A}\), \(\mathrm{tr}(\mathbf{A})\) denotes its trace, while \(\mathbf{A} \succeq \mathbf{0}\) means that \(\mathbf{A}\) is positive semidefinite. 
For an arbitrary-size matrix \(\mathbf{A}\), \(\mathrm{rank}(\mathbf{A})\), \(\mathbf{A}^T\), and \(\mathbf{A}^H\) denote its rank, transpose, and conjugate transpose, respectively, and \([\mathbf{A}]_{:,n}\) denotes its \(n\)-th column. 
For a vector \(\mathbf{a}\), \(\|\mathbf{a}\|\) denotes its Euclidean norm, and \([\mathbf{a}]_{m:n}\) denotes its \(m\)-th to \(n\)-th elements. For a complex number \(a\), \(|a|\) and \(\angle a\) denote its magnitude and phase, respectively.
\(\mathrm{diag}(a_1, \dots, a_N)\) denotes a diagonal matrix whose diagonal entries are \(a_1, \dots, a_N\).
\(\mathbf{0}_{M\times N}\) and \(\mathbf{1}_{M\times N}\) denote the all-zero and all-ones matrices with dimension \(M \times N\), respectively. \(\mathbb{C}^{M \times N}\) denotes the space of \(M \times N\) complex matrices. \(\mathbb{E}[\cdot]\) denotes the statistical expectation. 
\(j = \sqrt{-1}\) denotes the imaginary unit.  
For a function \(f(\mathbf{a})\), \(\nabla f(\mathbf{a})\) and \(\nabla^2 f(\mathbf{a})\) denote its gradient vector and Hessian matrix, respectively.

\section{System Model}

\begin{figure}[tb]
	\centering {\includegraphics[width=0.39\textwidth]{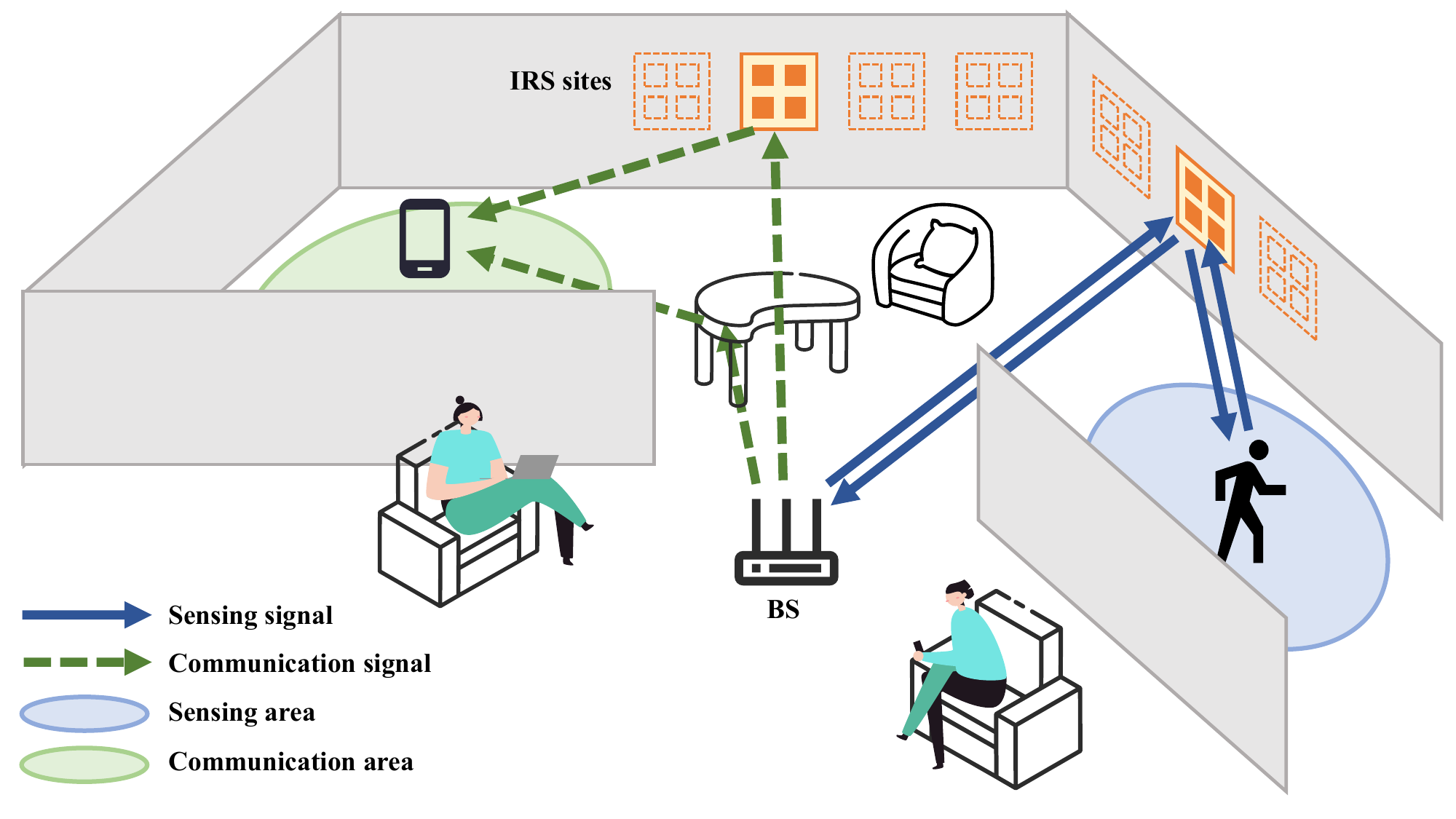}} 
	\caption{Illustration of the considered IRS-enabled ISAC system, in which IRSs need to be strategically deployed to facilitate sensing and communications at designated areas.}
\vspace{-0.2 cm}\end{figure}

As illustrated in Fig. 1, we consider an IRS-enabled ISAC system, in which multiple IRSs are strategically deployed to assist a BS to sense potential targets in a designated sensing area and deliver information to potential users in a designated communication area. It is assumed that the BS is equipped with a uniform linear array (ULA) of \(N_t\) antennas.
For analytical tractability, we adopt a discrete sampling approach, based on which the interested sensing and communication areas are represented by finite sets of \(P\) sensing points (SPs) and \(Q\) communication points (CPs), denoted by \(\mathcal{P} = \{1,\dots,P\}\) and \(\mathcal{Q} = \{1,\dots,Q\}\), respectively.\footnote{Notice that \(\mathcal{P}\) and \(\mathcal{Q}\) can be practically determined based on the prior measurement of potential sensing and communication requirements, and may exhibit spatial overlap in practical scenarios.} 
We denote the three-dimensional coordinates of the BS, each SP \(p\in\mathcal{P}\), and each CP \(q\in\mathcal{Q}\) by \(\mathbf{l}_0\), \(\mathbf{l}_p^\text{s}\), and \(\mathbf{l}_q^\text{c} \in \mathbb{R}^{3\times1}\), respectively. 
It is assumed that LoS propagation paths from the BS to the sensing and communication areas are blocked due to obstacles in the environment.
Such blockage necessitates the deployment of IRSs to enhance the sensing and communication performances. 

For ease of practical implementation, we consider a set of \(K\) candidate sites for IRS deployment, denoted by \(\mathcal{K} = \{1,\dots,K\}\), each with coordinate \(\mathbf{l}_k \in \mathbb{R}^{3\times1}\) denoting the IRS center.
Each deployed IRS consists of a set \(\mathcal{M} = \{1, \dots, M\}\) of \(M\) passive reflecting elements. 
We use a binary indicator vector \(\boldsymbol{\beta} = [\beta_1, \dots, \beta_K]^T\) to represent the IRS deployment decision, where \(\beta_k = 1\) indicates that an IRS should be deployed at candidate position \(k\in\mathcal{K}\), and \(\beta_k = 0\) otherwise.\footnote{While this paper considers fixed size and orientation for each IRS, the proposed IRS deployment framework can be extended to incorporate the IRS size and orientation optimization. This extension would involve treating different IRS sizes and orientations as distinct deployment options within the indicator vector \(\boldsymbol{\beta}\), while introducing additional constraints to ensure the uniqueness of IRS deployment at each candidate site.}


\subsection{Sensing and Communication Models}


The performance of IRS-enabled ISAC systems not only depends on the deployment of IRSs, but also on the BS's transmit beamforming and IRSs' reflective beamforming during operation. 
Specifically, the BS employs the active transmit beamforming, in which the transmit signal is denoted by \(\mathbf{x} \in \mathbb{C}^{N\times 1}\) with mean zero and covariance matrix \(\mathbf{R} = \mathbb{E}[\mathbf{x} \mathbf{x}^H]\) to be designed. The transmit power at the BS is denoted by \(P_0 = \mathrm{tr}(\mathbf{R})\), which is constrained by the maximum power budget \(\bar{P}_0\) such that \(P_0 \le \bar{P}_0\).
Simultaneously, each potentially deployed IRS at candidate site \(k \in \mathcal{K}\) implements the passive reflective beamforming by employing a diagonal phase-shift matrix \(\mathbf{\Theta}_k = \mathrm{diag}\big([e^{j\theta_{k,1}},\dots,e^{j\theta_{k,M}}]^T\big)\), where \(\theta_{k,m} \in [-\pi,\pi]\) denotes the phase shift of each reflecting element \(m\in\mathcal{M}\).

For wireless communication, we denote the channel matrix from the BS to IRS \(k\in\mathcal{K}\) as \(\mathbf{H}_{0, k} \in \mathbb{C}^{M\times N_t}\), the channel vector from IRS \(k\) to CP \(q \in \mathcal{Q}\) as \(\mathbf{h}_{k, q}^H \in \mathbb{C}^{1\times M}\), and the channel vector from the BS directly to CP \(q\) as \(\mathbf{h}_{0, q}^H \in \mathbb{C}^{1\times N_t}\), respectively.
It is assumed that the BS is able to acquire the CSI of these channels via CKM, as will be detailed in Section II-B. 
Under given IRS deployment indicator \(\boldsymbol{\beta}\), the received communication signal at each CP \(q\) is given by
\vspace{-0.1 cm}\begin{equation}
	y_{q}^\text{c} = \Big(\sum_{k\in\mathcal{K}} \beta_k \mathbf{h}_{k,q}^H \mathbf{\Theta}_k \mathbf{H}_{0,k} + \mathbf{h}_{0,q}^H\Big) \mathbf{x} + z_\text{c},
\vspace{-0.1 cm}\end{equation}
where \(z_\text{c}\) denotes the additive white Gaussian noise (AWGN) at the potential CU, which is a circularly symmetric complex Gaussian (CSCG) random variable with zero mean and variance \(\sigma_\text{c}^2\). 
We consider the received SNR at each CP \(q\) as the communication performance metric, which is given by
\vspace{-0.1 cm}\begin{equation} \label{gamcp}
	\begin{aligned}
		\gamma_{q}\big(\boldsymbol{\beta},& \{\mathbf{\Theta}_k\}, \mathbf{R}\big) = \frac{1}{\sigma_\text{c}^2} \Big(\sum_{k\in\mathcal{K}} \beta_k \mathbf{h}_{k,q}^H \mathbf{\Theta}_k \mathbf{H}_{0,k} + \mathbf{h}_{0,q}^H\Big) \mathbf{R} \\
		& \cdot \Big(\sum_{k\in\mathcal{K}} \beta_k \mathbf{h}_{k,q}^H \mathbf{\Theta}_k \mathbf{H}_{0,k} + \mathbf{h}_{0,q}^H\Big)^H.
	\end{aligned}
\vspace{-0.1 cm}\end{equation}

Different from wireless communication that exploits combined LoS and NLoS signal paths, the IRS-enabled sensing relies on the virtual LoS link toward the SPs.
Specifically, we denote the LoS channel from IRS \(k\in\mathcal{K}\) to SP \(p \in \mathcal{P}\) as \(\mathbf{g}_{k, p}^H \in \mathbb{C}^{1\times M}\).
The signal arriving at each SP \(p\) from the BS through the IRSs' LoS path (by omitting the noise and interference) is accordingly given by\footnote{While conventional sensing systems typically treat NLoS components as cluttering interference, our proposed framework leverages CKM to identify LoS paths and eliminate NLoS components from the received sensing signal, enabling the isolation of virtual LoS sensing channels.}
\vspace{-0.1 cm}\begin{equation} \label{y_p}
		y_{p}^\text{s} = \sum_{k\in\mathcal{K}} \beta_k \mathbf{g}_{k,p}^H \mathbf{\Theta}_k \mathbf{H}_{0,k} \mathbf{x}.
\vspace{-0.1 cm}\end{equation}
In this case, we adopt the illumination power of \(y_{p}^\text{s}\) at each SP \(p\) as the performance metric for sensing \cite{sankar2024beamforming, chepuri2023integrated}, which is given by
\vspace{-0.1 cm}\begin{equation} \label{gamsp}
	\begin{aligned}
		\rho_{p}\big(\boldsymbol{\beta},& \{\mathbf{\Theta}_k\}, \mathbf{R}\big) = \\ &\Big(\sum_{k\in\mathcal{K}} \beta_k \mathbf{g}_{k,p}^H \mathbf{\Theta}_k \mathbf{H}_{0,k}\Big) \mathbf{R} \Big(\sum_{k\in\mathcal{K}} \beta_k \mathbf{g}_{k,p}^H \mathbf{\Theta}_k \mathbf{H}_{0,k}\Big)^H.
	\end{aligned}
\vspace{-0.1 cm}\end{equation}
This metric effectively captures the useful signal power arriving at the target location through the virtual LoS paths created by IRSs, which is particularly suitable for various sensing applications such as intrusion detection and target localization \cite{sankar2024beamforming, chepuri2023integrated}.

\vspace{-0.1 cm}\subsection{Problem Formulation}

We jointly optimize the IRS deployment indicator \(\boldsymbol{\beta}\), the IRS reflective beamforming \(\{\mathbf{\Theta}_k\}\), and the BS transmit beamforming \(\mathbf{R}\) to minimize the system cost, while guaranteeing both sensing illumination power and communication SNR requirements for coverage. 
The total cost of our proposed system comprises two components, i.e., the capital cost that is proportional to the number of deployed IRSs, and the operational cost that is proportional to the BS power consumption.
Accordingly, the system cost is expressed as the weighted sum combination of the two components, i.e.,
\vspace{-0.1 cm}\begin{equation} \label{cost}
	\mathcal{C}(\boldsymbol{\beta}, P_0) = w_1 \mathbf{1}^T \boldsymbol{\beta} + w_2 P_0,
\vspace{-0.1 cm}\end{equation}
where \(w_1\) and \(w_2\) are the non-negative weighting coefficients for capital and operational costs, respectively.

Furthermore, we consider both transmit beamforming at the BS and reflective beamforming at the IRSs in our formulation. For transmit beamforming at BS, we adopt a real-time beamforming design, where the transmit covariance is optimized based on the specific CSI realization of SPs and CPs.\footnote{In our considered coverage optimization problem, we suppose that the transmit and reflective beamforming can be specifically designed for each SP and CP, by ignoring the case when multiple sensing targets and communication users coexist. This is reasonable for our study in the deployment phase.}
For reflective beamforming at IRSs, we consider the following two cases, since the adjustment of IRS reflective beamforming is usually less flexible as compared to the transmit beamforming at BS due to practical circuit restrictions \cite{feng2023reconfigurable}.

\begin{itemize}
	\item \textit{Case I with Quasi-stationary Reflective Beamforming at IRSs:} The IRS reflective beamforming is configured once at the beginning of the operation period and remains fixed throughout the whole period. 
	
	\item \textit{Case II with Dynamic Reflective Beamforming at IRSs:} The IRS reconfiguration is implemented dynamically at the same time-scale as the transmit beamforming at BS based on the real-time CSI of SPs and CPs. 
\end{itemize}
For both cases, we minimize the system cost \(\mathcal{C}(\boldsymbol{\beta}, P_0)\) while ensuring that the sensing illumination power at each SP and the communication SNR at each CP are no less than thresholds \(P_\text{s}\) and \(\Gamma_\text{c}\), respectively. 

For case I, the quasi-stationary implementation leads to a two-stage optimization as follows. In the first stage, the BS dynamically optimizes its transmit covariance matrix \(\mathbf{R}\) in real-time to maximize either the sensing illumination power \(\rho_{p}\big(\boldsymbol{\beta}, \{\mathbf{\Theta}_k\}, \mathbf{R}\big)\) at each SP \(p\) or the communication SNR \(\gamma_{q}\big(\boldsymbol{\beta}, \{\mathbf{\Theta}_k\}, \mathbf{R}\big)\) at each CP \(q\), while in the second stage, the system jointly optimizes the quasi-stationary parameters including IRS deployment indicator \(\boldsymbol{\beta}\), reflective beamforming \(\{\mathbf{\Theta}_k\}\), and BS transmit power \(P_0\).
The corresponding optimization problem is formulated as
\begin{subequations}
	\begin{align}
		\text{(P1):} &\ \min_{\boldsymbol{\beta}, \{\mathbf{\Theta}_k\}, P_0} \ w_1 \mathbf{1}^T \boldsymbol{\beta} + w_2 P_0\\
		\text{s.t.} &\ \max_{\mathbf{R}\succeq \mathbf{0}, \mathrm{tr}(\mathbf{R}) = P_0} \rho_{p}\big(\boldsymbol{\beta}, \{\mathbf{\Theta}_k\}, \mathbf{R}\big) \ge P_\text{s}, \forall p \in \mathcal{P}, \label{1s}\\
		&\ \max_{\mathbf{R}\succeq \mathbf{0}, \mathrm{tr}(\mathbf{R}) = P_0} \gamma_{q}\big(\boldsymbol{\beta}, \{\mathbf{\Theta}_k\}, \mathbf{R}\big) \ge \Gamma_\text{c}, \forall q \in \mathcal{Q}, \label{1c}\\
		&\ \beta_k \in \{0,1\}, \forall k \in \mathcal{K}, \label{1beta} \\
		&\ \mathbf{\Theta}_k = \mathrm{diag}\big([e^{j\theta_{k,1}},\dots,e^{j\theta_{k,M}}]^T\big), \nonumber \\ &\ \qquad \theta_{k,m} \in [-\pi,\pi], \forall k \in \mathcal{K}, m\in\mathcal{M}, \label{1Theta} \\
		&\ 0 \le P_0 \le \bar{P}_0 \label{1P0}.
	\end{align}
\end{subequations}
Besides sensing and communication coverage requirements \eqref{1s} and \eqref{1c}, problem (P1) is further constrained by structural restrictions \eqref{1beta} and \eqref{1Theta} on IRS deployment indicator and reflective beamforming matrices, respectively, as well as the maximum transmit power limitation \eqref{1P0}.

For case II, both BS transmission and IRS reconfiguration can be dynamically optimized for performance enhancement. As such, different from (P1) in which the reflective beamforming matrices \(\{\mathbf{\Theta}_k\}\) need to be remain unchanged over different SP and CP locations, in this case we can optimize \(\{\mathbf{\Theta}_k\}\) individually for each SP/CP to maximize the sensing illumination power in \eqref{1s} or the communication SNR in \eqref{1c}, thus providing a theoretical performance upper bound for our proposed framework. In this case, the system cost minimization problem is formulated as
\begin{subequations}
	\begin{align}
		\text{(P2):} &\ \min_{\boldsymbol{\beta}, P_0} \ w_1 \mathbf{1}^T \boldsymbol{\beta} + w_2 P_0\\
		\text{s.t.} &\ \max_{\{\mathbf{\Theta}_k\in\mathcal{V}_k\}, \mathbf{R}\succeq \mathbf{0}, \mathrm{tr}(\mathbf{R}) = P_0} \rho_{p}\big(\boldsymbol{\beta}, \{\mathbf{\Theta}_k\}, \mathbf{R}\big) \ge P_\text{s}, \forall p \in \mathcal{P}, \label{2s}\\
		&\ \max_{\{\mathbf{\Theta}_k\in\mathcal{V}_k\}, \mathbf{R}\succeq \mathbf{0}, \mathrm{tr}(\mathbf{R}) = P_0} \gamma_{q}\big(\boldsymbol{\beta}, \{\mathbf{\Theta}_k\}, \mathbf{R}\big) \ge \Gamma_\text{c}, \forall q \in \mathcal{Q}, \label{2c}\\
		&\ \mathcal{V}_k = \Big\{\mathbf{\Theta}_k \big| \mathbf{\Theta}_k = \mathrm{diag}\big([e^{j\theta_{k,1}},\dots,e^{j\theta_{k,M}}]^T\big), \nonumber \\ &\ \qquad \qquad \quad \theta_{k,m} \in [-\pi,\pi],  m\in\mathcal{M} \Big\}, \forall k \in \mathcal{K},\\
		&\ \eqref{1beta}, \eqref{1P0}. \nonumber
	\end{align}
\end{subequations}

Intuitively, both sensing and communication channels are unknown at the deployment stage, which poses difficulties for the joint design in problems (P1) and (P2). 
Another challenge in solving these non-convex mixed-integer optimization problems lies in the inherent coupling between the IRS deployment decisions and the reflective beamforming optimization. 
Compared to problem (P1), (P2) is even more challenging to solve due to the higher variable dimension of dynamic IRS reflective beamforming. 


\begin{figure}[tb]
	\centering 
	\subfigure[Environment model.] {\includegraphics[width=0.15\textwidth]{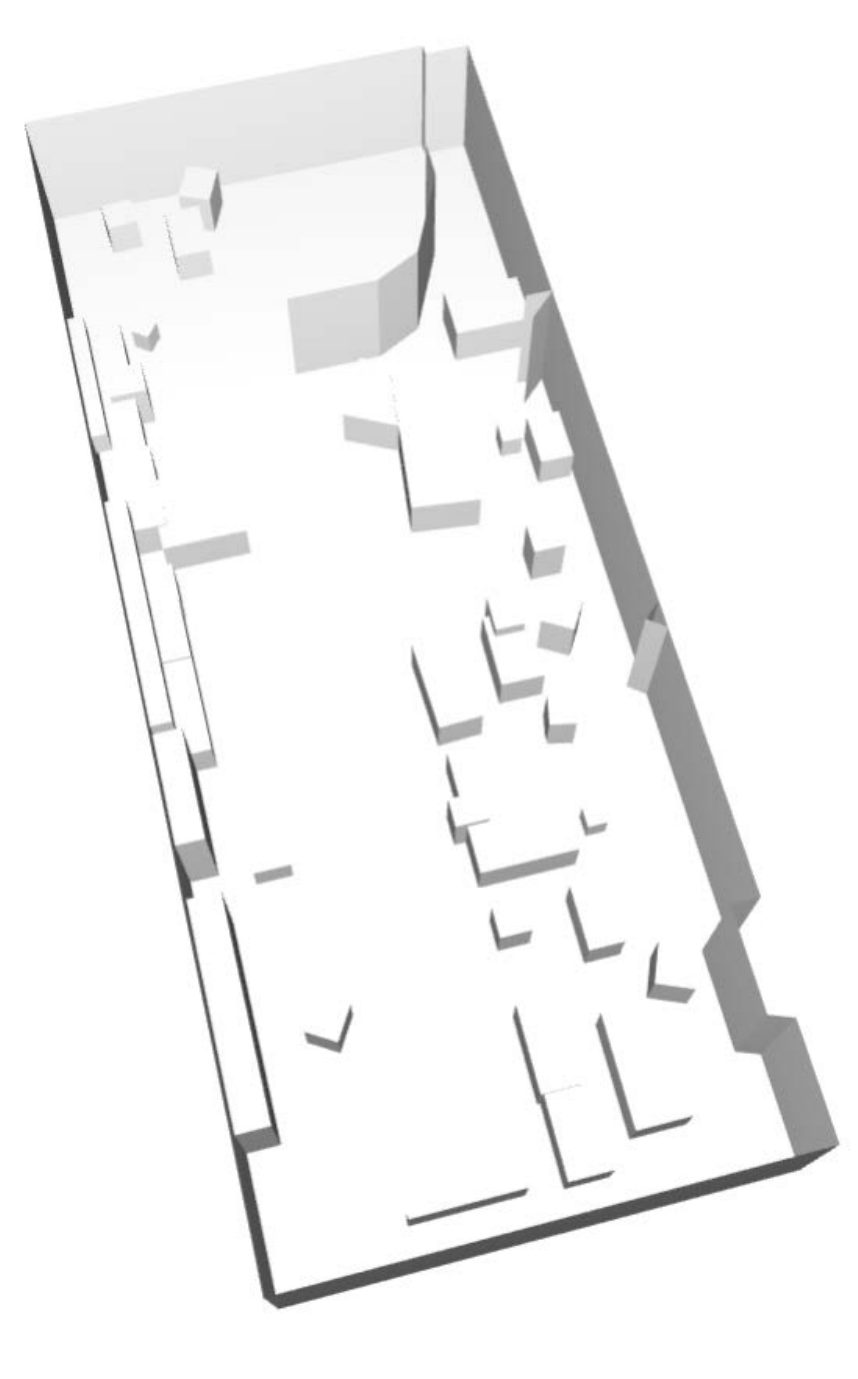}}
	\subfigure[Ray-tracing.] {\includegraphics[width=0.17\textwidth]{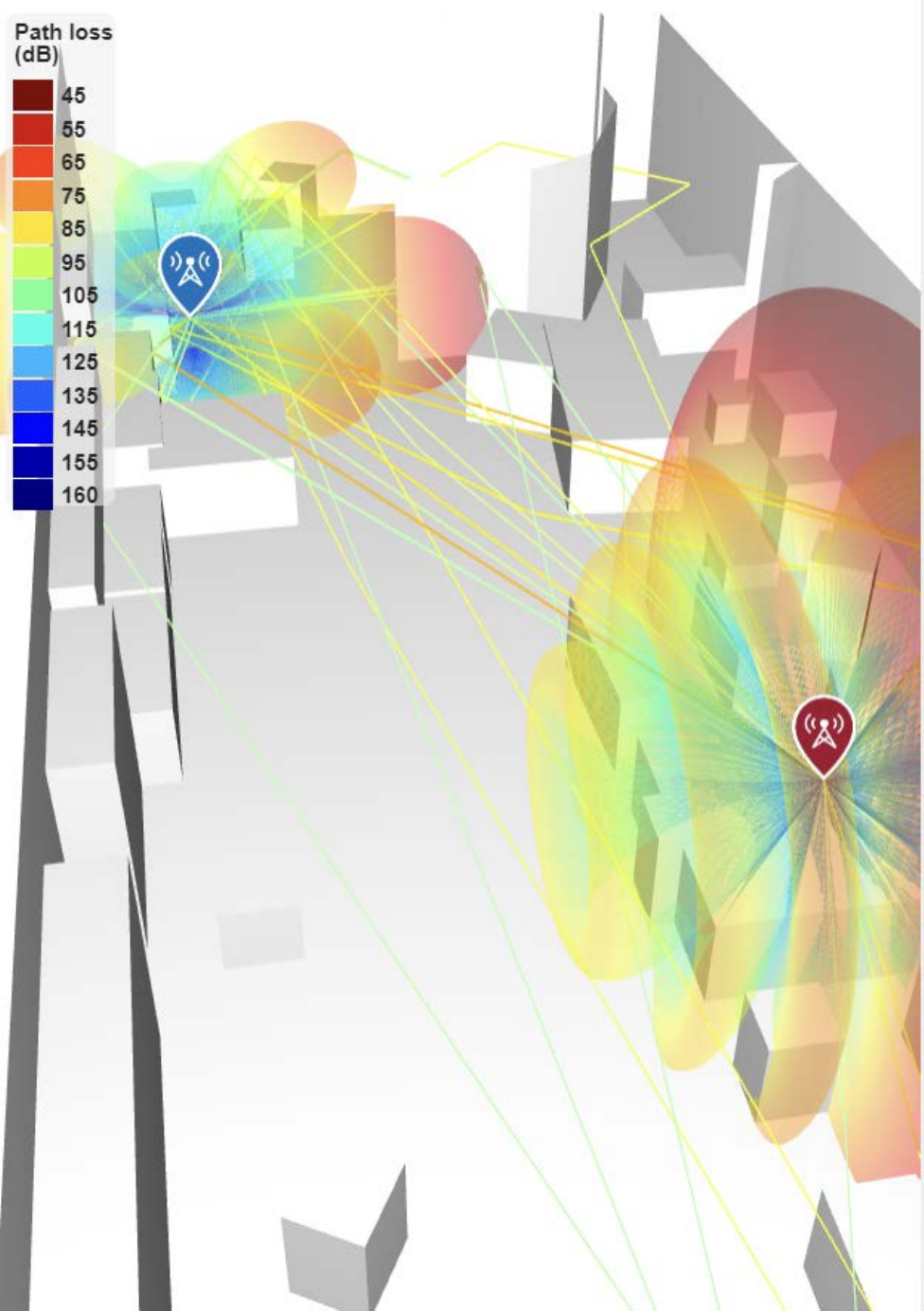}}
	\subfigure[Path loss map.] {\includegraphics[width=0.14\textwidth]{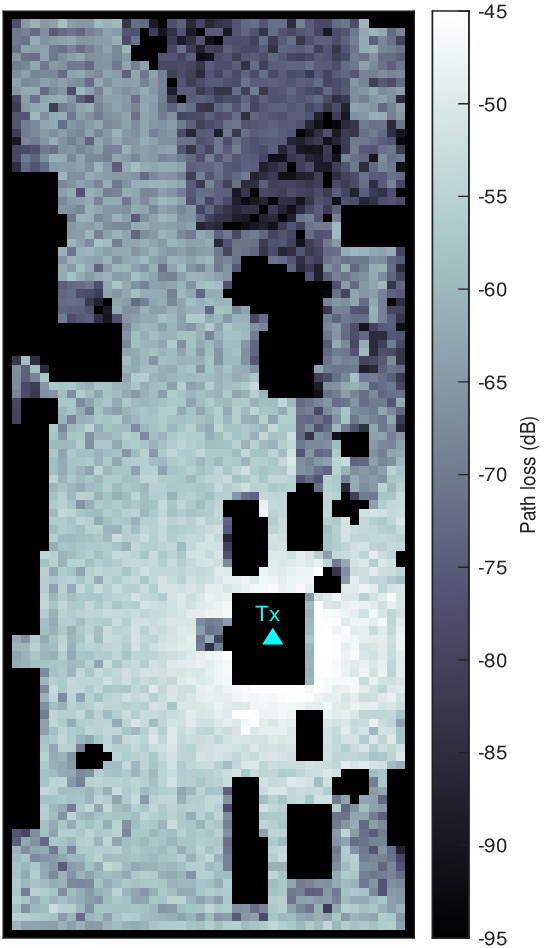}}
	\caption{An example of CKM construction by exploiting ray-tracing.}
	\label{Fig_rt}
\end{figure}

\textit{Remark: Exploiting CKM to Acquire CSI in the Deployment Stage.}
The optimization of problems (P1) and (P2) necessitates accurate acquisition of multi-path channels \(\{\mathbf{H}_{0,k}\}\), \(\{\mathbf{h}_{k,q}^H\}\), and \(\{\mathbf{h}_{0,q}^H\}\) for communication, as well as the LoS channel \(\{\mathbf{g}_{k,p}^H\}\) for sensing. However, the direct measurement at all SPs and CPs is impractical, especially without the actual deployment of IRSs at candidate locations. 
To tackle this issue, we leverage CKM \cite{zeng2021toward, zeng2024tutorial}, an environment-aware database tagged with location-specific channel knowledge, to acquire the CSI at transceiver and IRS sites during the deployment stage. 
As illustrated in Fig. \ref{Fig_rt}, CKM can be constructed by utilizing radio propagation modeling-based methods such as ray-tracing \cite{fuschini2015ray} and channel measurement at sparse reference locations.
To be specific, given the locations of BS \(\mathbf{l}_0\), candidate IRSs \(\{\mathbf{l}_k\}\), SPs \(\{\mathbf{l}_p^\text{s}\}\), and CPs \(\{\mathbf{l}_q^\text{c}\}\), as well as the spatial distribution of the environmental obstacles, we can obtain the multi-path channel knowledge between any two points from CKM.
For the communication channels between each candidate IRS location and BS or CPs, by exploiting, e.g., the ray-tracing technique, we can identify multi-path channel knowledge including the number of paths, as well as path loss, phase shift, and azimuth/elevation AoD/AoA for each path. Based on these knowledge, the CSI for communication channels \(\{\mathbf{H}_{0,k}\}\), \(\{\mathbf{h}_{k,q}^H\}\), and \(\{\mathbf{h}_{0,q}^H\}\) can be acquired.
For sensing, we additionally leverage the channel knowledge on the type of each path when acquiring the virtual LoS channels \(\{\mathbf{g}_{k,p}^H\}\) from candidate IRS locations to SPs.
This is different from communications and enables selective utilization of the LoS path while excluding NLoS components that would otherwise introduce clutter interference.

\section{SCA-based Relax-and-round Algorithm for Solving Problem (P1)}

In this section, we propose an SCA-based relax-and-round algorithm to solve problem (P1).
To this end, we first determine the optimal transmit covariance matrix \(\mathbf{R}\) with \(\mathrm{trace}(\mathbf{R}) = P_0\) to maximize \(\rho_{p}\big(\boldsymbol{\beta}, \{\mathbf{\Theta}_k\}, \mathbf{R}\big)\) and \(\gamma_{q}\big(\boldsymbol{\beta}, \{\mathbf{\Theta}_k\}, \mathbf{R}\big)\) in constraints \eqref{1s} and \eqref{1c}, respectively, \(\forall p\in\mathcal{P}, q\in\mathcal{Q}\). 
It is clear that 
the optimal \(\mathbf{R}\) for maximizing \(\rho_{p}\big(\boldsymbol{\beta}, \{\mathbf{\Theta}_k\}, \mathbf{R}\big)\) and \(\gamma_{q}\big(\boldsymbol{\beta}, \{\mathbf{\Theta}_k\}, \mathbf{R}\big)\) are given by
\begin{equation} \label{Rsc}
	\begin{aligned}
		\mathbf{R}_p^\text{s} &= \frac{P_0}{\|\mathbf{u}_p^\text{s}\|^2} \mathbf{u}_p^\text{s} (\mathbf{u}_p^\text{s})^H, \mathbf{u}_p^\text{s} = \sum_{k\in\mathcal{K}} \beta_k \mathbf{H}_{0,k}^H \mathbf{\Theta}_k^H \mathbf{g}_{k,p}, \\
		\mathbf{R}_q^\text{c} &= \frac{P_0}{\|\mathbf{u}_q^\text{c}\|^2} \mathbf{u}_q^\text{c} (\mathbf{u}_q^\text{c})^H, \mathbf{u}_q^\text{c} = \sum_{k\in\mathcal{K}} \beta_k \mathbf{H}_{0,k}^H \mathbf{\Theta}_k^H \mathbf{h}_{k,q} + \mathbf{h}_{0,q},
	\end{aligned}
\end{equation}
respectively.
Substituting \(\mathbf{R}_p^\text{s}\) and \(\mathbf{R}_q^\text{c}\) into constraints \eqref{1s} and \eqref{1c}, problem (P1) is equivalently reformulated as
\begin{equation} \label{p101}
	\begin{aligned}
		\min_{\boldsymbol{\beta}, \{\mathbf{\Theta}_k\}, P_0} &\ w_1 \mathbf{1}^T \boldsymbol{\beta} + w_2 P_0 \\
		\text{s.t.} &\ \Big\|\sum_{k\in\mathcal{K}} \beta_k \mathbf{H}_{0,k}^H \mathbf{\Theta}_k^H \mathbf{g}_{k,p}\Big\|^2 \ge \frac{P_\text{s}}{P_0}, \forall p\in\mathcal{P}, \\
		&\ \Big\|\sum_{k\in\mathcal{K}} \beta_k \mathbf{H}_{0,k}^H \mathbf{\Theta}_k^H \mathbf{h}_{k,q} + \mathbf{h}_{0,q}\Big\|^2  \ge \frac{\sigma_\text{c}^2 \Gamma_\text{c}}{P_0}, \forall q\in\mathcal{Q}, \\
		&\ \eqref{1beta}, \eqref{1Theta}, \eqref{1P0}.
	\end{aligned}
\end{equation}

For ease of notation, we define \(\mathbf{v}_k = [e^{j\theta_{k,1}}, \dots, e^{j\theta_{k,M}}]^H\) as the IRS reflective beamforming vector, \(\forall k \in \mathcal{K}\), and introduce auxiliary variable \(\mathbf{v} = [\mathbf{v}_1^T, \dots, \mathbf{v}_K^T]^T\). Additionally, we rearrange channel vectors \(\mathbf{g}_{k,p}\) and \(\mathbf{h}_{k,q}\) as \(\mathbf{G}_{k,p} = \mathrm{diag}(\mathbf{g}_{k,p})\) and \(\mathbf{H}_{k,q} = \mathrm{diag}(\mathbf{h}_{k,q})\), respectively. 
Accordingly, we rewrite problem \eqref{p101} as
\begin{subequations}
	\begin{align}
		\text{(P1.1):} &\ \min_{\boldsymbol{\beta}, \mathbf{v}, P_0} \ w_1 \mathbf{1}^T \boldsymbol{\beta} + w_2 P_0 \\
		\text{s.t.} &\ \Big\|\sum_{k\in\mathcal{K}} \beta_k \mathbf{H}_{0,k}^H \mathbf{G}_{k,p} \mathbf{v}_k\Big\|^2 \ge \frac{P_\text{s}}{P_0}, \forall p\in\mathcal{P}, \label{1.1S} \\
		&\ \Big\|\sum_{k\in\mathcal{K}} \beta_k \mathbf{H}_{0,k}^H \mathbf{H}_{k,q} \mathbf{v}_k + \mathbf{h}_{0,q}\Big\|^2  \ge \frac{\sigma_\text{c}^2 \Gamma_\text{c}}{P_0}, \forall q\in\mathcal{Q}, \label{1.1C} \\
		&\ \big|[\mathbf{v}]_i\big| = 1, \forall i \in \{1,\dots,KM\}, \label{1.1v} \\
		&\ \eqref{1beta}, \eqref{1P0}. \nonumber
	\end{align}
\end{subequations}

In the following, we propose an SCA-based relax-and-round algorithm to solve problem (P1.1). To be specific, we first relax constraint \eqref{1beta} and utilize the SCA technique to find a relaxed solution for \(\boldsymbol{\beta}\). We then round it back to the binary form and determine the corresponding solutions for \(\mathbf{v}\) and \(P_0\).

\subsection{Relaxation of Problem (P1.1)}

First, we relax the binary constraint \eqref{1beta} on IRS deployment indicator \(\boldsymbol{\beta}\) as \(0 \le \beta_k \le 1, \forall k \in \mathcal{K}\).
In addition, to facilitate the optimization of \(\boldsymbol{\beta}\), we provisionally relax the unit modulus constraint \eqref{1.1v} on IRS reflective beamforming \(\mathbf{v}\) as \(\big|[\mathbf{v}]_i\big| \le 1, \forall i \in \{1,\dots,KM\}\). 
Consequently, problem (P1.1) is relaxed as
\begin{subequations}
	\begin{align}
		\text{(P1.2):} &\ \min_{\boldsymbol{\beta}, \mathbf{v}, 0\le P_0\le \bar{P}_0} \ w_1 \mathbf{1}^T \boldsymbol{\beta}  + w_2 P_0 \\
		\text{s.t.} &\ f_p^\text{s}(\mathbf{v}, \boldsymbol{\beta}) + \frac{P_\text{s}}{P_0} \le 0, \forall p\in\mathcal{P}, \label{3.2s}\\
		&\ f_q^\text{c}(\mathbf{v}, \boldsymbol{\beta}) + \frac{\sigma_\text{c}^2 \Gamma_\text{c}}{P_0} \le 0, \forall q\in\mathcal{Q}, \label{3.2c}\\
		&\ 0 \le \beta_k \le 1, \forall k \in \mathcal{K}, \label{3.2beta}\\
		&\ \big|[\mathbf{v}]_i\big| \le 1, \forall i \in \{1,\dots,KM\}, \label{3.2v}
	\end{align}
\end{subequations}
where \(f_p^\text{s}(\mathbf{v}, \boldsymbol{\beta})\) and \(f_q^\text{c}(\mathbf{v}, \boldsymbol{\beta})\) are respectively defined as
\begin{equation} \label{fsc}
	\begin{aligned}
		f_p^\text{s}(\mathbf{v}, \boldsymbol{\beta}) &= - \Big\|\sum_{k\in\mathcal{K}} \beta_k \mathbf{H}_{0,k}^H \mathbf{G}_{k,p} \mathbf{v}_k\Big\|^2, \\
		f_q^\text{c}(\mathbf{v}, \boldsymbol{\beta}) &= - \Big\|\sum_{k\in\mathcal{K}} \beta_k \mathbf{H}_{0,k}^H \mathbf{H}_{k,q} \mathbf{v}_k + \mathbf{h}_{0,q}\Big\|^2,
	\end{aligned}
\end{equation}
for all \(p\in\mathcal{P}, q\in\mathcal{Q}\).

\subsection{SCA-based Solution for Problem (P1.2)}

In this subsection, we utilize the SCA technique to solve problem (P1.2), aiming to determine the relaxed \(\boldsymbol{\beta}\).
First, to address the non-convexity in \(f_p^\text{s}(\mathbf{v}, \boldsymbol{\beta})\) and \(f_q^\text{c}(\mathbf{v}, \boldsymbol{\beta})\), we utilize the SCA technique to approximate them into a series of convex forms. Specifically, in each iteration \(t\) of the SCA algorithm, we denote the current feasible point of \(\mathbf{v}\) and \(\boldsymbol{\beta}\) as \(\mathbf{v}^{(t)}\) and \(\boldsymbol{\beta}^{(t)}\), respectively. Notice that since \(f_p^\text{s}(\mathbf{v}, \boldsymbol{\beta})\) and \(f_q^\text{c}(\mathbf{v}, \boldsymbol{\beta})\) are neither convex nor concave, we cannot directly derive linear bounds for them. Instead, at each current point \(\mathbf{v}^{(t)}\) and \(\boldsymbol{\beta}^{(t)}\), we approximate \(f_p^\text{s}(\mathbf{v}, \boldsymbol{\beta})\) and \(f_q^\text{c}(\mathbf{v}, \boldsymbol{\beta})\) as the following quadratic bounds \cite{yang2017unified, yang2018successive}.
\begin{equation} \label{fspcp}
	\begin{aligned}
		&f_p^\text{s}(\mathbf{v}, \boldsymbol{\beta}) \le \tilde{f}_p^\text{s}\big(\mathbf{v}, \boldsymbol{\beta} | \mathbf{v}^{(t)}, \boldsymbol{\beta}^{(t)}\big) \triangleq f_p^\text{s}(\mathbf{v}^{(t)}, \boldsymbol{\beta}^{(t)}) \\ &+ \Re \left\{ \nabla f_p^\text{s}(\mathbf{v}^{(t)}, \boldsymbol{\beta}^{(t)})^H \begin{bmatrix}
			\mathbf{v} - \mathbf{v}^{(t)} \\ \boldsymbol{\beta} - \boldsymbol{\beta}^{(t)}
		\end{bmatrix} \right\} + \frac{\mu_p^\text{s}}{2} \left\|\begin{bmatrix}
			\mathbf{v} - \mathbf{v}^{(t)} \\ \boldsymbol{\beta} - \boldsymbol{\beta}^{(t)}
		\end{bmatrix}\right\|^2, \\
		&f_q^\text{c}(\mathbf{v}, \boldsymbol{\beta}) \le \tilde{f}_q^\text{c}\big(\mathbf{v}, \boldsymbol{\beta} | \mathbf{v}^{(t)}, \boldsymbol{\beta}^{(t)}\big) \triangleq f_q^\text{c}(\mathbf{v}^{(t)}, \boldsymbol{\beta}^{(t)}) \\ &+ \Re \left\{ \nabla f_q^\text{c}(\mathbf{v}^{(t)}, \boldsymbol{\beta}^{(t)})^H \begin{bmatrix}
			\mathbf{v} - \mathbf{v}^{(t)} \\ \boldsymbol{\beta} - \boldsymbol{\beta}^{(t)}
		\end{bmatrix} \right\} + \frac{\mu_q^\text{c}}{2} \left\|\begin{bmatrix}
			\mathbf{v} - \mathbf{v}^{(t)} \\ \boldsymbol{\beta} - \boldsymbol{\beta}^{(t)}
		\end{bmatrix}\right\|^2.
	\end{aligned}
\end{equation}
In \eqref{fspcp}, the gradients of \(f_p^\text{s}(\mathbf{v}, \boldsymbol{\beta})\) and \(f_q^\text{c}(\mathbf{v}, \boldsymbol{\beta})\) with respect to (w.r.t.) \([\mathbf{v}^T, \boldsymbol{\beta}^T]^T\) are given by
\begin{equation}
	\begin{aligned}
		\nabla & f_p^\text{s}(\mathbf{v}, \boldsymbol{\beta}) = -2 \big(\mathbf{U}_p^\text{s}(\mathbf{v}, \boldsymbol{\beta})\big)^H \sum_{k\in\mathcal{K}} \beta_k \mathbf{H}_{0,k}^H \mathbf{G}_{k,p} \mathbf{v}_k, \\
		\nabla & f_q^\text{c}(\mathbf{v}, \boldsymbol{\beta}) = -2 \big(\mathbf{U}_q^\text{c}(\mathbf{v}, \boldsymbol{\beta})\big)^H \Big(\sum_{k\in\mathcal{K}} \beta_k \mathbf{H}_{0,k}^H \mathbf{H}_{k,q} \mathbf{v}_k + \mathbf{h}_{0,q}\Big),
	\end{aligned}
\end{equation}
respectively, in which we have
\begin{equation}
	\begin{aligned}
		\mathbf{U}_p^\text{s}(\mathbf{v}, \boldsymbol{\beta}) = \big[&\beta_1 \mathbf{H}_{0,1}^H \mathbf{G}_{1,p}, \dots, \beta_K \mathbf{H}_{0,K}^H \mathbf{G}_{K,p}, \\ & \mathbf{H}_{0,1}^H \mathbf{G}_{1,p} \mathbf{v}_1, \dots, \mathbf{H}_{0,K}^H \mathbf{G}_{K,p} \mathbf{v}_K\big], \\
		\mathbf{U}_q^\text{c}(\mathbf{v}, \boldsymbol{\beta}) = \big[&\beta_1 \mathbf{H}_{0,1}^H \mathbf{H}_{1,q}, \dots, \beta_K \mathbf{H}_{0,K}^H \mathbf{H}_{K,q}, \\ & \mathbf{H}_{0,1}^H \mathbf{H}_{1,q} \mathbf{v}_1, \dots, \mathbf{H}_{0,K}^H \mathbf{H}_{K,q} \mathbf{v}_K\big].
	\end{aligned}
\end{equation}
In addition, based on the descent lemma \cite{bertsekas1999nonlinear}, we can empirically set hyper-parameters \(\mu_p^\text{s}\) and \(\mu_q^\text{c}\) in \eqref{fspcp} as sufficiently large values that satisfy \(\mu_p^\text{s} \ge L_p^\text{s}\) and \(\mu_q^\text{c} \ge L_q^\text{c}\) for ensuring convergence \cite{yang2017unified, yang2018successive}, where \(L_p^\text{s}\) and \(L_q^\text{c}\) denote the Lipschitz constants for \(f_p^\text{s}(\mathbf{v}, \boldsymbol{\beta})\) and \(f_q^\text{c}(\mathbf{v}, \boldsymbol{\beta})\), respectively.\footnote{According to the descent lemma \cite{bertsekas1999nonlinear}, \(L_p^\text{s}\) and \(L_q^\text{c}\) satisfy \(L_p^\text{s} \ge \|\nabla^2 f_p^\text{s}(\mathbf{v}, \boldsymbol{\beta})\|\) and \(L_q^\text{c} \ge \|\nabla^2 f_q^\text{c}(\mathbf{v}, \boldsymbol{\beta})\|\), respectively, \(\forall \mathbf{v}, \boldsymbol{\beta}\). As such, \(\mu_p^\text{s}\) and \(\mu_q^\text{c}\) can be approximated according to the upper bound of the Frobenius norms of \(\nabla^2 f_p^\text{s}(\mathbf{v}, \boldsymbol{\beta})\) and \(\nabla^2 f_q^\text{c}(\mathbf{v}, \boldsymbol{\beta})\), where we omit the details for brevity.} Accordingly, at the current point \(\mathbf{v}^{(t)}\) and \(\boldsymbol{\beta}^{(t)}\), we approximate problem (P1.2) as
\begin{equation} \label{3SCA}
	\begin{aligned}
		\min_{\boldsymbol{\beta}, \mathbf{v}, 0\le P_0\le \bar{P}_0} &\ w_1 \mathbf{1}^T \boldsymbol{\beta}  + w_2 P_0 \\
		\text{s.t.} &\ \tilde{f}_p^\text{s}\big(\mathbf{v}, \boldsymbol{\beta} | \mathbf{v}^{(t)}, \boldsymbol{\beta}^{(t)}\big) + \frac{P_\text{s}}{P_0} \le 0, \forall p\in\mathcal{P}, \\
		&\ \tilde{f}_q^\text{c}\big(\mathbf{v}, \boldsymbol{\beta} | \mathbf{v}^{(t)}, \boldsymbol{\beta}^{(t)}\big) + \frac{\sigma_\text{c}^2 \Gamma_\text{c}}{P_0} \le 0, \forall q\in\mathcal{Q}, \\
		&\ \eqref{3.2beta}, \eqref{3.2v}.
	\end{aligned}
\end{equation}

Problem \eqref{3SCA} is a quadratically constrained quadratic program (QCQP), which can be efficiently solved by standard convex optimization tools such as CVX \cite{cvx2012}. The obtained optimal solution of \(\mathbf{v}\) and \(\boldsymbol{\beta}\) in each iteration \(t\) is then updated as the local point of the next iteration, i.e., \(\mathbf{v}^{(t+1)}\) and \(\boldsymbol{\beta}^{(t+1)}\).
The iteration continues until a prescribed convergence criterion is met, and we denote the converged solution of \(\boldsymbol{\beta}\) to problem (P1.2) as \(\boldsymbol{\beta}^\text{SCA}\).
Notably, as problem \eqref{3SCA} consistently yields non-increasing objective values over iterations and the total system cost function is lower bounded, the convergence of the SCA-based iterative algorithm is ensured.

\subsection{Construction of Binary IRS Deployment Indicator}


After obtaining the relaxed IRS deployment indicator solution \(\boldsymbol{\beta}^\text{SCA}\) to (P1.2), the remaining challenge in solving problem (P1.1) is to round them back to the binary IRS deployment indicators, and determine the corresponding solutions for \(\mathbf{v}\) and \(P_0\). In the following, we introduce a greedy search-based rounding process.

To start with, we consider the optimization of problem (P1.1) over \(\mathbf{v}\) and \(P_0\) with given \(\boldsymbol{\beta}\), which is expressed as
\begin{equation*}
	\begin{aligned}
		\text{(P1.3):} &\ \min_{\mathbf{v}, 0 \le P_0 \le \bar{P}_0} \ P_0 \\
		\text{s.t.} &\ \eqref{1.1S}, \eqref{1.1C}, \eqref{1.1v}.
	\end{aligned}
\end{equation*}
To facilitate discussion, we temporarily assume that the solutions of \(\mathbf{v}\) and \(P_0\) to problem (P1.3) with given \(\boldsymbol{\beta}\) are obtained as \(\mathbf{v}^\star(\boldsymbol{\beta})\) and \(P_0^\star(\boldsymbol{\beta})\), respectively, which will be specified later.

Moreover, we sort the non-zero elements of \(\boldsymbol{\beta}^\text{SCA}\) in ascending order, and denote the corresponding indices as \(\Xi \triangleq \{\xi_1, \dots, \xi_{\hat{K}}\}\), where \(\hat{K}\) is the number of non-zero elements in \(\boldsymbol{\beta}^\text{SCA}\), and \(\beta_{\xi_i}^\text{SCA}\) denotes the \(i\)-th smallest non-zero element.
We then initialize \(\boldsymbol{\beta}\) by setting the elements corresponding to the non-zero ones in \(\boldsymbol{\beta}^\text{SCA}\) to one, i.e.,
\begin{equation} \label{betak}
	\beta_k = \begin{cases}
		0, & \beta_k^\text{SCA} = 0 \\
		1, & \text{otherwise}
	\end{cases}, \forall k \in \mathcal{K}.
\end{equation}

In each iteration \(i \in \{1,\dots,\hat{K}\}\) of the rounding process, we temporarily set the element in \(\boldsymbol{\beta}\) that corresponds to the \(i\)-th smallest non-zero one in \(\boldsymbol{\beta}^\text{SCA}\) to zero, i.e., \(\beta_{\xi_i} = 0\). We then check the feasibility of problem (P1.3) with the current \(\boldsymbol{\beta}\). Specifically, if problem (P1.3) is feasible or equivalently \(P_0^\star(\boldsymbol{\beta}) \le \bar{P}_0\) holds, we record the current \(\boldsymbol{\beta}\) as a candidate solution to problem (P1.1), denoted by \(\boldsymbol{\beta}^\star\).
Accordingly, the corresponding solutions of \(\mathbf{v}\) and \(P_0\) are given by \(\mathbf{v}^\star(\boldsymbol{\beta}^\star)\) and \(P_0^\star(\boldsymbol{\beta}^\star)\), respectively.
Otherwise, if problem (P1.3) is infeasible, we restore \(\boldsymbol{\beta}\) by setting \(\beta_{\xi_i} = 1\). Then, we proceed to the next iteration. Among all candidate solutions, the one that minimizes the objective \(\mathcal{C}^\star = w_1 \mathbf{1}^T \boldsymbol{\beta}^\star + w_2 P_0^\star(\boldsymbol{\beta}^\star)\) is selected as the final solution to problem (P1.1), 
denoted by \(\boldsymbol{\beta}^\text{I}\), \(\mathbf{v}^\text{I}\), and \(P_0^\text{I}\), respectively.

Now, we introduce the solution of problem (P1.3) for determining \(\mathbf{v}^\star(\boldsymbol{\beta})\) and \(P_0^\star(\boldsymbol{\beta})\) with given \(\boldsymbol{\beta}\).
By introducing the auxiliary variable \(\bar{\mathbf{v}} = [\mathbf{v}^T, 1]^T\), problem (P1.3) is reformulated as
\begin{subequations} \label{Pv1}
	\begin{align}
		\min_{\bar{\mathbf{v}}, 0 \le P_0 \le \bar{P}_0} &\ P_0 \\
		\text{s.t.} &\ \big\|\mathbf{F}_p^\text{s}(\boldsymbol{\beta}) \bar{\mathbf{v}}\big\|^2 \ge \frac{P_\text{s}}{P_0}, \forall p\in\mathcal{P}, \label{v1S}\\
		&\ \big\|\mathbf{F}_q^\text{c}(\boldsymbol{\beta}) \bar{\mathbf{v}}\big\|^2 \ge \frac{\sigma_\text{c}^2 \Gamma_\text{c}}{P_0}, \forall q\in\mathcal{Q}, \label{v1C}\\
		&\ \big|[\bar{\mathbf{v}}]_i\big| = 1, \forall i \in \{1,\dots,KM\}, \\ 
		&\ [\bar{\mathbf{v}}]_{KM+1} = 1,
	\end{align}
\end{subequations}
where \(\mathbf{F}_p^\text{s}(\boldsymbol{\beta})\) and \(\mathbf{F}_q^\text{c}(\boldsymbol{\beta})\) are defined as
\begin{equation} \label{Fsc}
	\begin{aligned}
		\mathbf{F}_p^\text{s}(\boldsymbol{\beta}) = \big[&\beta_1 \mathbf{H}_{0,1}^H \mathbf{G}_{1,p}, \dots, \beta_K \mathbf{H}_{0,K}^H \mathbf{G}_{K,p}, \mathbf{0}_{N\times1}\big], \\
		\mathbf{F}_q^\text{c}(\boldsymbol{\beta}) = \big[&\beta_1 \mathbf{H}_{0,1}^H \mathbf{H}_{1,q}, \dots, \beta_K \mathbf{H}_{0,K}^H \mathbf{H}_{K,q}, \mathbf{h}_{0,q}\big],
	\end{aligned}
\end{equation}
respectively. 

Notice that problem \eqref{Pv1} is non-convex. To solve it, we introduce an auxiliary variable \(\bar{\mathbf{V}} = \bar{\mathbf{v}} \bar{\mathbf{v}}^H\), and rewrite \eqref{Pv1} as
\begin{subequations} \label{PvSDR}
	\begin{align}
		\text{(P1.4):} &\ \min_{\bar{\mathbf{V}} \succeq \mathbf{0}, 0 \le P_0 \le \bar{P}_0} \ P_0\\
		\text{s.t.} &\ \mathrm{tr}\Big(\big(\mathbf{F}_p^\text{s}(\boldsymbol{\beta})\big)^H \mathbf{F}_p^\text{s}(\boldsymbol{\beta}) \bar{\mathbf{V}}\Big) \ge \frac{P_\text{s}}{P_0}, \forall p\in\mathcal{P}, \\
		&\ \mathrm{tr}\Big(\big(\mathbf{F}_q^\text{c}(\boldsymbol{\beta})\big)^H \mathbf{F}_q^\text{c}(\boldsymbol{\beta}) \bar{\mathbf{V}}\Big) \ge \frac{\sigma_\text{c}^2 \Gamma_\text{c}}{P_0}, \forall q\in\mathcal{Q}, \\		 
		&\ \big[\bar{\mathbf{V}}\big]_{i,i} = 1, \forall i\in\{1, \dots, KM+1\}, \\
		&\ \mathrm{rank}(\bar{\mathbf{V}}) = 1. \label{vSDRone}
	\end{align}
\end{subequations}
Problem (P1.4) is still non-convex due to the rank-one constraint \eqref{vSDRone}. To address this, we utilize the semidefinite relaxation (SDR) technique to relax constraint \eqref{vSDRone}, and denote the relaxed problem as (SDR1.4). Problem (SDR1.4) is a semidefinite program (SDP), which can be efficiently solved by standard convex optimization tools. We denote its optimal solution of \(\bar{\mathbf{V}}\) as \(\bar{\mathbf{V}}^*\). 

Based on \(\bar{\mathbf{V}}^*\), we adopt Gaussian randomization to generate a feasible rank-one solution for the original problem \eqref{Pv1}. To be specific, we first generate multiple random realizations \(\tilde{\mathbf{v}}\) as CSCG random vectors with zero mean and covariance \(\bar{\mathbf{V}}^*\). Accordingly, we construct normalized candidate solutions as \(\tilde{\mathbf{v}}^* = e^{j \angle \frac{\tilde{\mathbf{v}}}{[\tilde{\mathbf{v}}]_{KM+1}}}\). The randomization is independently performed multiple times, and the objective value achieved by each randomization \(\tilde{\mathbf{v}}^*\) is recorded as \(P_0^*(\tilde{\mathbf{v}}^*) = \max_{p\in\mathcal{P}, q\in\mathcal{Q}} \left\{\frac{P_\text{s}}{\|\mathbf{F}_p^\text{s}(\boldsymbol{\beta}) \tilde{\mathbf{v}}^*\|^2}, \frac{\sigma_\text{c}^2 \Gamma_\text{c}}{\|\mathbf{F}_q^\text{c}(\boldsymbol{\beta}) \tilde{\mathbf{v}}^*\|^2}\right\}\).
Then, the solution of \(P_0\) to problem \eqref{Pv1}, or equivalently (P1.3), is approximated as the minimum one achieved by the randomizations, which is expressed as 
\begin{equation}
	P_0^\star(\boldsymbol{\beta}) = \min_{\tilde{\mathbf{v}}^*} P_0^*(\tilde{\mathbf{v}}^*).
\end{equation}
Notice that to guarantee the feasibility, the obtained solution should satisfy \(P_0^\star(\boldsymbol{\beta}) \le \bar{P}_0\).
The corresponding randomization is selected as the solution of \(\bar{\mathbf{v}}\) for problem \eqref{Pv1}, i.e., \(\bar{\mathbf{v}}^\star(\boldsymbol{\beta}) = \arg \min_{\tilde{\mathbf{v}}^*} P_0^*(\tilde{\mathbf{v}}^*)\). Accordingly, the solution of \(\mathbf{v}\) to problem (P1.3) is given by
\begin{equation}
	\mathbf{v}^\star(\boldsymbol{\beta}) = \big[\bar{\mathbf{v}}^\star(\boldsymbol{\beta})\big]_{1:KM}.
\end{equation}

\subsection{Complete Algorithm for Solving Problem (P1.1)}


Now, we summarize the SCA-based relax-and-round algorithm for solving problem (P1.1).
We first determine the initial point of \(\boldsymbol{\beta}\) and \(\mathbf{v}\) for the relaxed problem (P1.2), denoted by \(\boldsymbol{\beta}^{(1)}\) and \(\mathbf{v}^{(1)}\), respectively. To be specific, we initialize \(\boldsymbol{\beta}\) by setting  \(\boldsymbol{\beta}^{(1)} = \mathbf{1}\), and initialize \(\mathbf{v}\) by solving problem (P1.3) with given \(\boldsymbol{\beta}^{(1)}\), i.e., \(\mathbf{v}^{(1)} = \mathbf{v}^\star(\boldsymbol{\beta}^{(1)})\).
Next, we utilize SCA to solve problem (P1.2). In each iteration \(t\) of the SCA algorithm, we solve problem \eqref{3SCA} at local point \(\mathbf{v}^{(t)}\) and \(\boldsymbol{\beta}^{(t)}\), and update the local point of the next iteration as \(\mathbf{v}^{(t+1)}\) and \(\boldsymbol{\beta}^{(t+1)}\). The operation is iterated until a prescribed convergence criterion is met. 
Finally, we round the converged solution \(\boldsymbol{\beta}^\text{SCA}\) for problem (P1.2) back to the binary solution \(\boldsymbol{\beta}^\text{I}\) for problem (P1.1), and accordingly determine the solutions of \(\mathbf{v}^\text{I}\) and \(P_0^\text{I}\).

The computational complexity of the SCA-based algorithm is dominated by the process of solving the relaxed problem (P1.2) in Section III-B and the rounding process in Section III-C.
Specifically, when solving the relaxed problem (P1.2), we denote the number of SCA iterations as \(\mathcal{I}_\text{SCA}\). In each iteration, with a given solution accuracy \(\epsilon > 0\), the worst-case complexity to solve the QCQP in \eqref{3SCA} is given by \(\mathcal{O}\big((P+Q)^{0.5} (KM)^3 \log\frac{1}{\epsilon}\big)\) \cite{ben2001lectures}.
Furthermore, during the rounding process, there are at most \(K\) iterations. In each iteration, with a given solution accuracy \(\epsilon > 0\), the worst-case complexity to solve the SDP in (SDR1.4) is given by \(\mathcal{O}\big(\max\{P+Q, KM\}^4 (KM)^{0.5} \log\frac{1}{\epsilon}\big)\) \cite{luo2010semidefinite}. Also, the complexity of the Gaussian randomization process is \(\mathcal{O}\big(N_\text{GR} (P+Q) (KM)^2\big)\), where \(N_\text{GR}\) denotes the number of random realizations.
Therefore, the total complexity for the SCA-based algorithm is approximated as
\begin{equation}
	\begin{aligned}
		\mathcal{T}_\text{I}^\text{SCA} =& \mathcal{O}\bigg(\mathcal{I}_\text{SCA} (P+Q)^{0.5} (KM)^3 \log\frac{1}{\epsilon} \\
		&+ K\Big(\max\{P+Q, KM\}^4 (KM)^{0.5} \log\frac{1}{\epsilon} \\
		&+ N_\text{GR} (P+Q) (KM)^2\Big) \bigg).
	\end{aligned}
\end{equation}

\section{SCA-based Relax-and-round Algorithm for Solving Problem (P2)}

In this section, we present the solution to problem (P2).
As problems (P1) and (P2) exhibit similar structures, we modify the SCA-based relax-and-round solution for problem (P1) to solve problem (P2).

First, we maximize \(\rho_{p}\big(\boldsymbol{\beta}, \{\mathbf{\Theta}_k\}, \mathbf{R}\big)\) and \(\gamma_{q}\big(\boldsymbol{\beta}, \{\mathbf{\Theta}_k\}, \mathbf{R}\big)\) over \(\mathbf{R}\) and \(\{\mathbf{\Theta}_k\}\) in constraints \eqref{2s} and \eqref{2c}, respectively, \(\forall p\in\mathcal{P}, q\in\mathcal{Q}\). Similar to problem (P1), the optimal \(\mathbf{R}\) is given by \(\mathbf{R}_p^\text{s}\) and \(\mathbf{R}_q^\text{c}\) in \eqref{Rsc}, respectively. As for the optimization of \(\{\mathbf{\Theta}_k\}\), we denote the IRS reflective beamforming that maximizes \(\rho_{p}\big(\boldsymbol{\beta}, \{\mathbf{\Theta}_k\}, \mathbf{R}\big)\) and \(\gamma_{q}\big(\boldsymbol{\beta}, \{\mathbf{\Theta}_k\}, \mathbf{R}\big)\) as \(\{\mathbf{\Theta}_{k,p}^\text{s}\}\) and \(\{\mathbf{\Theta}_{k,q}^\text{c}\}\), respectively. For ease of notation, we introduce auxiliary variables \(\mathbf{v}_p^\text{s} = \big[(\mathbf{v}_{1,p}^\text{s})^T, \dots, (\mathbf{v}_{K,p}^\text{s})^T\big]^T\) and \(\mathbf{v}_q^\text{c} = \big[(\mathbf{v}_{1,q}^\text{c})^T, \dots, (\mathbf{v}_{K,q}^\text{c})^T\big]^T\), where \(\mathbf{v}_{k,p}^\text{s}\) and \( \mathbf{v}_{k,q}^\text{c}\) satisfy that \((\mathbf{\Theta}_{k,p}^\text{s})^H = \mathrm{diag}(\mathbf{v}_{k,p}^\text{s})\) and \((\mathbf{\Theta}_{k,q}^\text{c})^H = \mathrm{diag}(\mathbf{v}_{k,q}^\text{c})\), respectively, \(\forall k\in\mathcal{K}\).
Consequently, problem (P2) is equivalently reformulated as
\begin{equation}
	\begin{aligned}
		\text{(P2.1):} &\ \min_{\boldsymbol{\beta}, \{\mathbf{v}_p^\text{s}\}, \{\mathbf{v}_q^\text{c}\}, P_0} \ w_1 \mathbf{1}^T \boldsymbol{\beta} + w_2 P_0 \\
		\text{s.t.} &\ \Big\|\sum_{k\in\mathcal{K}} \beta_k \mathbf{H}_{0,k}^H \mathbf{G}_{k,p} \mathbf{v}_{k,p}^\text{s}\Big\|^2 \ge \frac{P_\text{s}}{P_0}, \forall p\in\mathcal{P}, \\
		&\ \Big\|\sum_{k\in\mathcal{K}} \beta_k \mathbf{H}_{0,k}^H \mathbf{H}_{k,q} \mathbf{v}_{k,q}^\text{c} + \mathbf{h}_{0,q}\Big\|^2 \ge \frac{\sigma_\text{c}^2 \Gamma_\text{c}}{P_0}, \forall q\in\mathcal{Q}, \\
		&\ \big|[\mathbf{v}_p^\text{s}]_i\big| = 1, \forall p\in\mathcal{P}, i \in \{1,\dots,KM\}, \\
		&\ \big|[\mathbf{v}_q^\text{c}]_i\big| = 1, \forall q\in\mathcal{Q}, i \in \{1,\dots,KM\}, \\
		&\ \eqref{1beta}, \eqref{1P0}.
	\end{aligned}
\end{equation}

Similar as that for solving problem (P1.1), we first relax the binary constraint \eqref{1beta} to \eqref{3.2beta}, and re-express problem (P2.1) as
\begin{equation*}
	\begin{aligned}
		\text{(P2.2):} &\ \min_{\boldsymbol{\beta}, \{\mathbf{v}_p^\text{s}\}, \{\mathbf{v}_q^\text{c}\}, 0\le P_0\le \bar{P}_0} \ w_1 \mathbf{1}^T \boldsymbol{\beta} + w_2 P_0 \\
		\text{s.t.} &\ f_p^\text{s}(\mathbf{v}_p^\text{s}, \boldsymbol{\beta}) + \frac{P_\text{s}}{P_0} \le 0, \forall p\in\mathcal{P}, \\
		&\ f_q^\text{c}(\mathbf{v}_q^\text{c}, \boldsymbol{\beta}) + \frac{\sigma_\text{c}^2 \Gamma_\text{c}}{P_0} \le 0, \forall q\in\mathcal{Q}, \\
		&\ \big|[\mathbf{v}_p^\text{s}]_i\big| \le 1, \forall p\in\mathcal{P}, i \in \{1,\dots,KM\}, \\
		&\ \big|[\mathbf{v}_q^\text{c}]_i\big| \le 1, \forall q\in\mathcal{Q}, i \in \{1,\dots,KM\}, \\
		&\ \eqref{3.2beta},
	\end{aligned}
\end{equation*}
where \(f_p^\text{s}(\mathbf{v}_p^\text{s}, \boldsymbol{\beta})\) and \(f_q^\text{c}(\mathbf{v}_q^\text{c}, \boldsymbol{\beta})\) are defined in \eqref{fsc} by replacing \(\mathbf{v}\) with \(\mathbf{v}_p^\text{s}\) and \(\mathbf{v}_q^\text{c}\), respectively. Problem (P2.2) can be solved similarly as problem (P1.2), by utilizing the SCA technique and iteratively solving an approximation problem like \eqref{3SCA}. 

Then, based on the relaxed solution for \(\boldsymbol{\beta}\), we modify the rounding process in Section III-C to recover \(\boldsymbol{\beta}\) to the binary form, and accordingly determine the solutions of \(\{\mathbf{v}_p^\text{s}\}\), \(\{\mathbf{v}_q^\text{c}\}\), and \(P_0\). The difference is that when optimizing problem (P2.1) over \(\{\mathbf{v}_p^\text{s}\}\), \(\{\mathbf{v}_q^\text{c}\}\), and \(P_0\) with given \(\boldsymbol{\beta}\), we equivalently solve the following \(P+Q\) sub-problems, each for minimizing the system cost while guaranteeing the sensing/communication performance at each SP \(p\in\mathcal{P}\) or CP \(q\in\mathcal{Q}\).
\begin{equation*}
	\begin{aligned}
		(\text{P2.3.}p): &\ \min_{\bar{\mathbf{v}}_p^\text{s}, 0\le P_0\le \bar{P}_0} \ P_0 \\
		\text{s.t.} &\ \big\|\mathbf{F}_p^\text{s}(\boldsymbol{\beta}) \bar{\mathbf{v}}_p^\text{s}\big\|^2 \ge \frac{P_\text{s}}{P_0}\\ 
		&\ \big|[\bar{\mathbf{v}}_p^\text{s}]_i\big| = 1, \forall i \in \{1, \dots, KM\}, \\
		&\ [\bar{\mathbf{v}}_p^\text{s}]_{KM+1} = 1,
	\end{aligned}
\end{equation*}
\begin{equation*}
	\begin{aligned}
		(\text{P2.4.}q): &\ \min_{\bar{\mathbf{v}}_q^\text{c}, 0\le P_0\le \bar{P}_0} \ P_0 \\
		\text{s.t.} &\ \big\|\mathbf{F}_q^\text{c}(\boldsymbol{\beta}) \bar{\mathbf{v}}_q^\text{c}\big\|^2 \ge \frac{\sigma_\text{c}^2 \Gamma_\text{c}}{P_0}\\ 
		&\ \big|[\bar{\mathbf{v}}_q^\text{c}]_i\big| = 1, \forall i \in \{1, \dots, KM\}, \\
		&\ [\bar{\mathbf{v}}_q^\text{c}]_{KM+1} = 1,
	\end{aligned}
\end{equation*}
where \(\bar{\mathbf{v}}_p^\text{s} = \big[(\mathbf{v}_p^\text{s})^T, 1\big]^T\) and \(\bar{\mathbf{v}}_q^\text{c} = \big[(\mathbf{v}_q^\text{c})^T, 1\big]^T\) are auxiliary variables defined similarly as \(\bar{\mathbf{v}}\).
Problems (P2.3) and (P2.4) can be solved similarly as problem \eqref{Pv1} by utilizing the SDR and Gaussian randomization techniques. We denote the obtained solution to sub-problem (P2.3.\(p\)) as \(\bar{\mathbf{v}}_p^{\text{s},\star\star}\) and \(P_{0,p}^{\star\star}\), and that to sub-problem (P2.4.\(q\)) as  \(\bar{\mathbf{v}}_q^{\text{c},\star\star}\) and \(P_{0,q}^{\star\star}\). During the rounding process, with given candidate solution of \(\boldsymbol{\beta}\), denoted as \(\boldsymbol{\beta}^{\star\star}\), we update the corresponding solutions of \(\{\mathbf{v}_p^\text{s}\}\), \(\{\mathbf{v}_q^\text{c}\}\), and \(P_0\) as \(\mathbf{v}_p^{\text{s},\star\star} = \big[\bar{\mathbf{v}}_p^{\text{s},\star\star}\big]_{1:KM}\), \(\mathbf{v}_q^{\text{c},\star\star} = \big[\bar{\mathbf{v}}_q^{\text{c},\star\star}\big]_{1:KM}\), and \(P_0^{\star\star} = \max_{p\in\mathcal{P},q\in\mathcal{Q}}\left\{P_{0,p}^{\star\star}, P_{0,q}^{\star\star}\right\}\), respectively.
Among all candidate solutions, the one that minimizes the objective \(\mathcal{C}^{\star\star} = w_1 \mathbf{1}^T \boldsymbol{\beta}^{\star\star} + w_2 P_0^{\star\star}\) is selected as the final solution to problem (P2.1), denoted by \(\boldsymbol{\beta}^\text{II}\), \(\{\mathbf{v}_p^{\text{s,II}}\}\), \(\{\mathbf{v}_q^{\text{c,II}}\}\) and \(P_0^\text{II}\), respectively.

Compared with the algorithm for solving (P1), the computational complexity of the SCA-based algorithm for (P2) differs in the rounding process. To be specific, compared to problem (P1.3), the worst-case complexity for solving problems (P2.3) and (P2.4) increases to \(\mathcal{O}\big((P+Q) (KM)^{4.5} \log\frac{1}{\epsilon} + N_\text{GR} (P+Q) (KM)^2\big)\) \cite{luo2010semidefinite}.
Therefore, the total complexity is approximated as
\begin{equation}
	\begin{aligned}
		\mathcal{T}_\text{II}^\text{SCA} &= \mathcal{O}\bigg(\mathcal{I}_\text{SCA} (P+Q)^{0.5} (KM)^3 \log\frac{1}{\epsilon} \\
		+& K\Big((P+Q) (KM)^{4.5} \log\frac{1}{\epsilon} + N_\text{GR} (P+Q) (KM)^2\Big) \bigg).
	\end{aligned}
\end{equation}

\section{Other Heuristic Solutions}

In this section, we propose a heuristic algorithm with CBD, which solves the joint IRS deployment and reflective beamforming optimization problems (P1) and (P2) with lower computational complexity. We also introduce a benchmark scheme with RRB for performance comparison.

\subsection{Channel-based Deployment (CBD)}

In this scheme, we solve problems (P1) and (P2) by directly determining the IRS deployment weights, i.e., the relaxed \(\boldsymbol{\beta}\), based on the channel condition between each candidate IRS location to SPs and CPs. As a single IRS cannot cover the whole sensing and communication areas, it is not appropriate to directly determine the deployment of each IRS based on its worst-case channel. Instead, we consider the weighted sum of the channel power gain from each IRS to all SPs and CPs.

To be specific, we define a contribution coefficient \(\eta_k\) for each candidate IRS \(k\in\mathcal{K}\) as the sum of its channel power gains toward the SPs and CPs, weighted by the corresponding sensing and communication requirements, i.e., 
\begin{equation}
	\eta_k = P_\text{s} \sum_{p\in\mathcal{P}} \|\mathbf{g}_{k,p}\|^2 + \sigma_\text{c}^2 \Gamma_\text{c} \sum_{q\in\mathcal{Q}} \|\mathbf{h}_{k,q}\|^2,
\end{equation}
where we normalize the communication SNR requirement \(\Gamma_\text{c}\) by the noise power \(\sigma_\text{c}^2\) in order to have a similar scale as the sensing illumination power requirement \(P_\text{s}\). 
Consequently, the IRS deployment weight \(\boldsymbol{\beta}^\text{CBD} = [\beta_1^\text{CBD}, \dots, \beta_K^\text{CBD}]^T\) is determined as
\begin{equation}
	\beta_k^\text{CBD} = \frac{\eta_k}{\max_{i\in\mathcal{K}} \eta_i}, \forall k\in\mathcal{K}.
\end{equation}

Based on the IRS deployment weight \(\boldsymbol{\beta}^\text{CBD}\), we then employ our proposed rounding processes in Sections III-C and IV for cases I and II to recover the binary IRS deployment indicators, and determine the corresponding solutions to IRS reflective beamforming and BS transmit power by solving problem (P1.3) and problems (P2.3) and (P2.4), respectively. 

Compared to the SCA-based algorithm, the CBD design directly determines the IRS deployment weights and skips the process of solving problems (P1.2) and (P2.2), and the time complexity accordingly reduces to
\begin{equation}
	\begin{aligned}
		\mathcal{T}_\text{I}^\text{CBD} =& \mathcal{O}\bigg(K\Big(\max\{P+Q, KM\}^4 (KM)^{0.5} \log\frac{1}{\epsilon} \\
		&+ N_\text{GR} (P+Q) (KM)^2\Big) \bigg), \\
		\mathcal{T}_\text{II}^\text{CBD} =& \mathcal{O}\bigg(K\Big((P+Q) (KM)^{4.5} \log\frac{1}{\epsilon} \\
		&+ N_\text{GR} (P+Q) (KM)^2\Big) \bigg),
	\end{aligned}
\end{equation}
for cases I and II, respectively.

\subsection{IRSs with Random Reflective Beamforming (RRB)}

For this scheme, we randomly determine the IRS reflective beamforming. Then, we optimize the IRS deployment indicator and BS transmit power with given IRS reflective beamforming.


To be specific, we randomly generate the IRS reflective beamforming \(\mathbf{v}\), by setting the phase of each element in \(\mathbf{v}\) to be uniformly distributed in \([-\pi,\pi]\). Then, with given \(\mathbf{v}\), we optimally determine the IRS deployment indicator \(\boldsymbol{\beta}\) and BS transmit power \(P_0\) by solving the following problem.
\begin{subequations} \label{Pb2}
	\begin{align}
			\min_{\boldsymbol{\beta}, \bar{\mathbf{B}}, 0 \le P_0 \le \bar{P}_0} &\ w_1 \mathrm{tr}(\mathbf{E}_0 \bar{\mathbf{B}}) + w_2 P_0 \\
			\text{s.t.} &\ \mathrm{tr}\Big(\big(\mathbf{D}_p^\text{s}(\mathbf{v})\big)^H \mathbf{D}_p^\text{s}(\mathbf{v}) \bar{\mathbf{B}}\Big) \ge \frac{P_\text{s}}{P_0}, \forall p \in \mathcal{P}, \label{b2S}\\
			&\ \mathrm{tr}\Big(\big(\mathbf{D}_q^\text{c}(\mathbf{v})\big)^H \mathbf{D}_q^\text{c}(\mathbf{v}) \bar{\mathbf{B}}\Big) \ge \frac{\sigma_\text{c}^2 \Gamma_\text{c}}{P_0}, \forall q \in \mathcal{Q}, \label{b2C}\\
			&\ \bar{\mathbf{B}} = [\boldsymbol{\beta}^T, 1]^T [\boldsymbol{\beta}^T, 1], \beta_k \in \{0,1\}, \forall k \in \mathcal{K}, \label{b2B}
		\end{align}
\end{subequations}
where we introduce the auxiliary variable \(\bar{\mathbf{B}}\) in \eqref{b2B} and define \(\mathbf{D}_p^\text{s}(\mathbf{v}) = [\mathbf{H}_{0,1}^H \mathbf{G}_{1,p} \mathbf{v}_1, \dots, \mathbf{H}_{0,K}^H \mathbf{G}_{K,p} \mathbf{v}_1, \mathbf{0}_{N\times1}]\), \(\mathbf{D}_q^\text{c}(\mathbf{v}) = [\mathbf{H}_{0,1}^H \mathbf{H}_{1,q} \mathbf{v}_1, \dots, \mathbf{H}_{0,K}^H \mathbf{H}_{K,q} \mathbf{v}_K, \mathbf{h}_{0,q}]\), and \(\mathbf{E}_0 = \begin{bmatrix}
	\mathbf{0}_{K\times K} & \mathbf{1}_{K\times1} \\ \mathbf{0}_{1\times K} & 0
\end{bmatrix}\), respectively.

Notice that problem \eqref{Pb2} is equivalent to
	\begin{subequations} \label{Pb3}
		\begin{align}
			\min_{\bar{\mathbf{B}}, 0 \le P_0 \le \bar{P}_0} &\ w_1 \mathrm{tr}(\mathbf{E}_0 \bar{\mathbf{B}}) + w_2 P_0\\
			\text{s.t.} &\ \eqref{b2S}, \eqref{b2C}, \\
			&\ [\bar{\mathbf{B}}]_{:,k} \le [\bar{\mathbf{B}}]_{:,K+1}, \forall k\in\mathcal{K}, \label{b3le} \\
			&\ \mathrm{tr}(\mathbf{E}_1 \bar{\mathbf{B}}) \ge \mathrm{tr}^2(\mathbf{E}_0 \bar{\mathbf{B}}), \label{b3ge} \\
			&\ [\bar{\mathbf{B}}]_{i,j} \in \{0,1\}, \forall i,j \in \mathcal{K}\cup\{K+1\}, \label{b3bin} \\
			&\ [\bar{\mathbf{B}}]_{K+1,K+1} = 1, \label{b3K+1} \\
			&\ \bar{\mathbf{B}} = \bar{\mathbf{B}}^T, \label{b3sym}
		\end{align}
	\end{subequations}
where we define \(\mathbf{E}_1 = \begin{bmatrix}
	\mathbf{1}_{K\times K} & \mathbf{0}_{K\times1} \\ \mathbf{0}_{1\times K} & 0
\end{bmatrix}\). This is because constraints \eqref{b3le} and \eqref{b3ge} guarantee that there are exactly \((\mathbf{1}^T\boldsymbol{\beta})^2\) elements being one in \([\bar{\mathbf{B}}]_{1:K,1:K}\). Therefore, together with structural constraints \eqref{b3bin}, \eqref{b3K+1}, and \eqref{b3sym}, they are equivalent to constraint \eqref{b2B}. Problem \eqref{Pb3} is a mixed-integer optimization problem with linear and second-order cone constraints, which can be directly solved by standard solvers such as Mosek \cite{mosek2025}. In this case, the computational complexity of solving \eqref{Pb3} is in exponential time w.r.t. \(K\), which is difficult to be precisely quantified.


\section{Numerical Results}

\begin{figure}[tb]
	\centering 
	\subfigure[Locations of BS, candidate IRSs, SPs and CPs.]
	{\includegraphics[width=0.8\columnwidth]{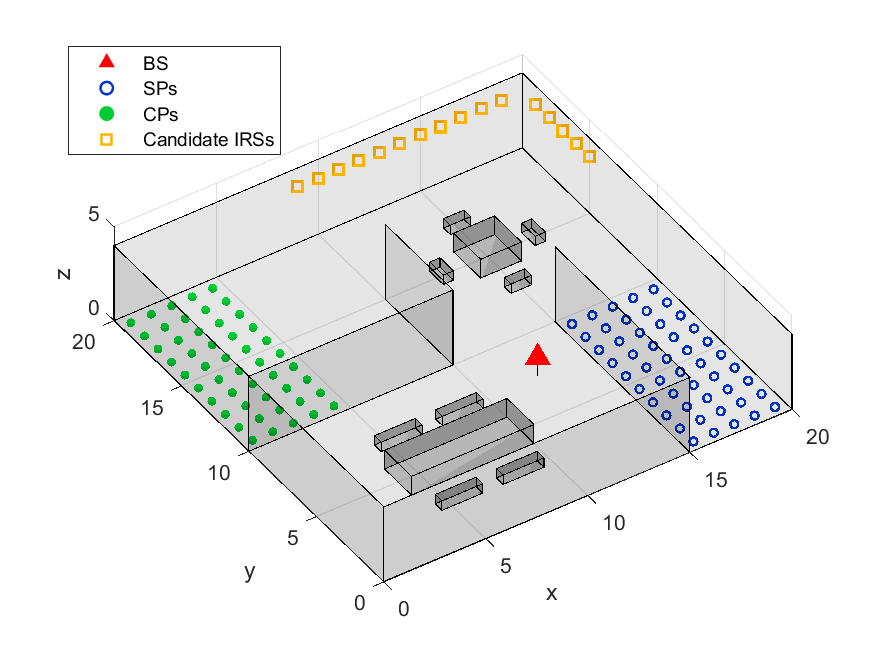}}
	\subfigure[Coverage over the SPs and CPs.]
	{\includegraphics[width=0.9\columnwidth]{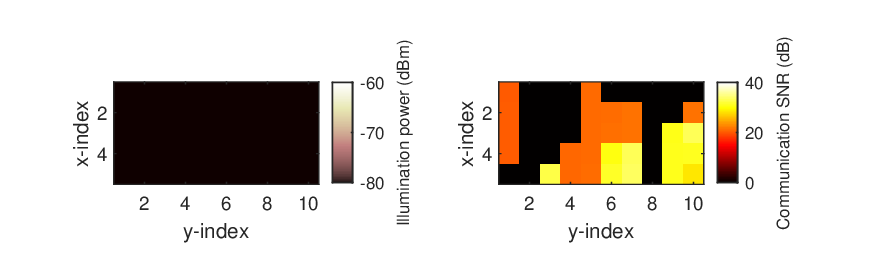}}
	\caption{Simulation scenario and sensing/communication coverage without deploying IRSs.}
	\label{fig_0}
\end{figure}
This section provides numerical results to evaluate the performance of our proposed joint IRS deployment and reflective beamforming designs. 
We consider a practical smart home scenario, in which the locations of the BS, candidate IRSs, SPs, and CPs are shown in Fig. \ref{fig_0}(a). The BS is deployed to guarantee the LoS coverage of the main rooms, while IRSs are deployed to achieve the sensing coverage to the entry (for, e.g, intrusion detection) and simultaneously achieve the communication coverage to a secondary room. For such a scenario, we utilize the ray-tracing function in the Communications Toolbox of MATLAB to generate the CKM. As shown in Fig. \ref{fig_0}(b), without deploying IRSs, there is no LoS link between the BS and the sensing area, leading to a zero sensing illumination power coverage at all SPs. On the other hand, although a portion of the communication area is covered with relatively high SNR, the BS still fails to communicate with some CPs (e.g., the communication SNR at some CPs is lower than \(0\) dB).

The following system parameters are considered unless specified otherwise. The BS is equipped with a ULA of \(N_t = 8\) transmit antennas. There are \(K = 16\) candidate IRS locations, and each IRS is equipped with \(M = 64\) reflecting elements. We consider a carrier frequency of 3.5 GHz, and the spaces between adjacent BS antennas and IRS reflecting elements are both set as half a wavelength. The sensing and communication areas are both discretized into \(P = Q = 50\) SPs and CPs. The maximum transmit power at the BS and the noise power at each CP are set as \(\bar{P}_0 = 30\) dBm and \(\sigma_\text{c}^2 = -80\) dBm, respectively \cite{fu2023active, kang2023double}. 
Moreover, without loss of generality, we normalize the IRS deployment cost weight as \(w_1 = 1\).

First, we demonstrate the convergence and complexity performances of our proposed designs.
\begin{figure}[tb]
	\centering {\includegraphics[width=0.4\textwidth]{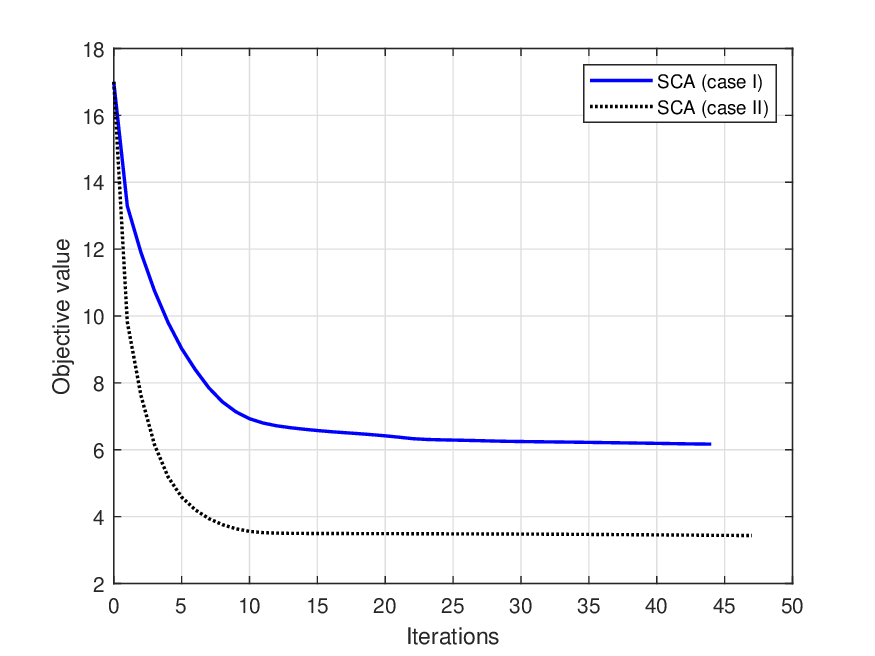}}
	\caption{Convergence performance of SCA-based algorithms.}
	\label{fig_iter}
\end{figure}
Fig. \ref{fig_iter} shows the convergence performance of our proposed SCA designs, with \(P_\text{s} = -74\) dBm, \(\Gamma_\text{c} = 6\) dB, and weight \(w_2 = 1\). We demonstrate the objective values of the SCA-based solution for relaxed problems (P1.2) and (P2.2) in cases I and II, respectively. It is observed that for both cases, the SCA-based solutions achieve non-increasing objectives and converge within a similar number of iterations. Furthermore, the gap between the two converged objective values reflects their different ability in reducing system cost, which will be demonstrated in the following results.

\begin{figure}[tb]
	\centering
	\subfigure[Execution time versus the number of candidate IRSs \(K\).]
	{\includegraphics[width=0.8\columnwidth]{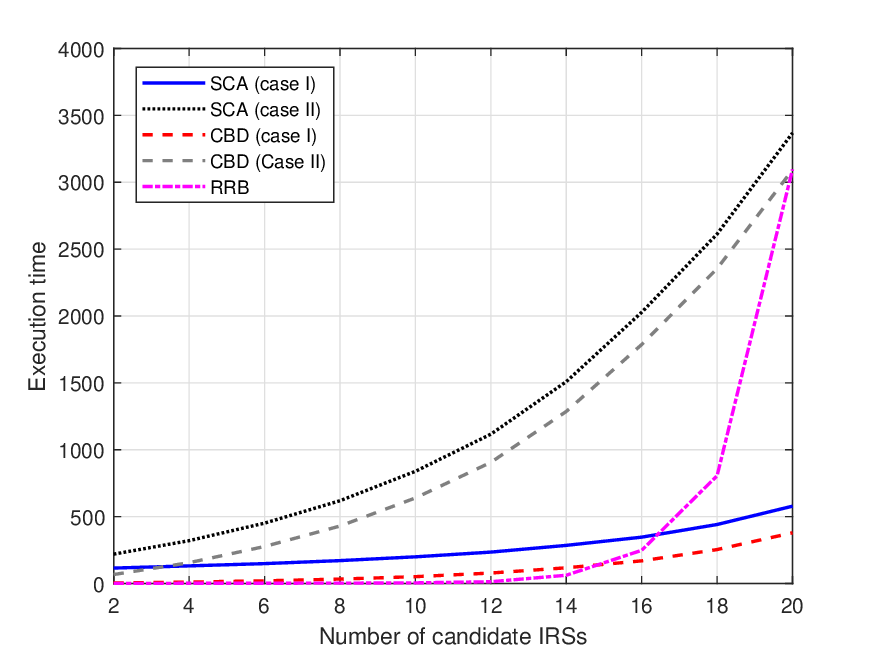}}
	\subfigure[Execution time versus the number of sample points \(P+Q\).]
	{\includegraphics[width=0.8\columnwidth]{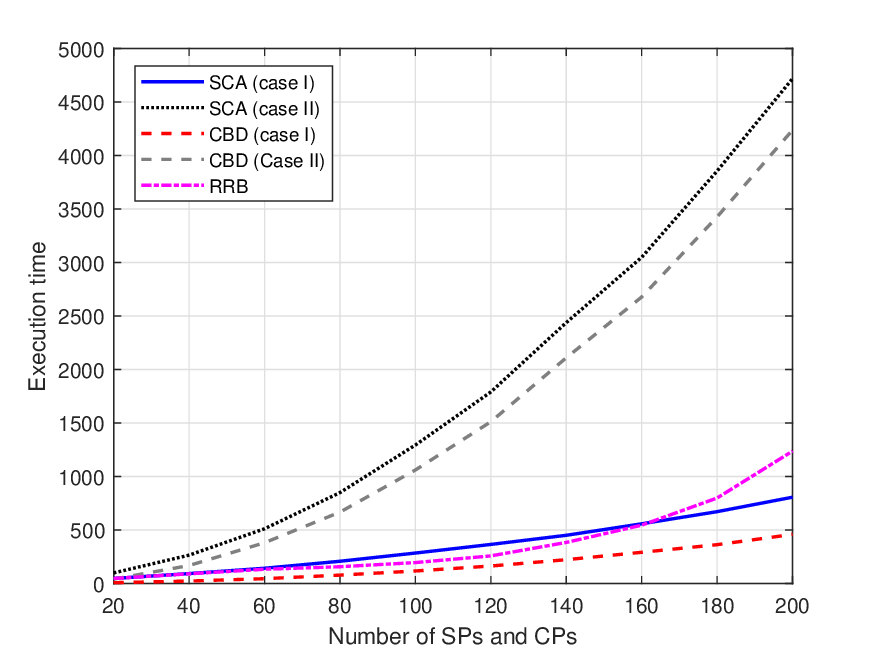}}
	\caption{Execution time performance of our proposed designs.}
	\label{fig_time}
\end{figure}
Fig. \ref{fig_time} shows the computational complexity of our proposed SCA, CBD, and RRB designs under different system setups. It is observed that in case I, the execution time of CBD is significantly lower than that of SCA. In this case, the complexity of SCA-based algorithm for determining IRS deployment weight is comparable to that of rounding process, and the CBD design skips the former procedure.
In case II, as the numbers of sample points \(P+Q\) and candidate IRSs \(K\) become large, the complexity of the rounding process increases much faster, as specified in Section IV. As a result, the execution time of SCA and CBD becomes comparable.
Furthermore, the execution time of the RRB design increases more drastically due to the exponential complexity for solving the mixed-integer problem \eqref{Pb3}, especially for the case when \(K\) is large.

Next, we evaluate the system cost achieved by our proposed designs under different setups.
\begin{figure}[tb]
	\centering 
	\subfigure[IRS deployment cost versus \(\Gamma_\text{c}\) with \(P_\text{s} = -74\) dBm.]
	{\includegraphics[width=0.8\columnwidth]{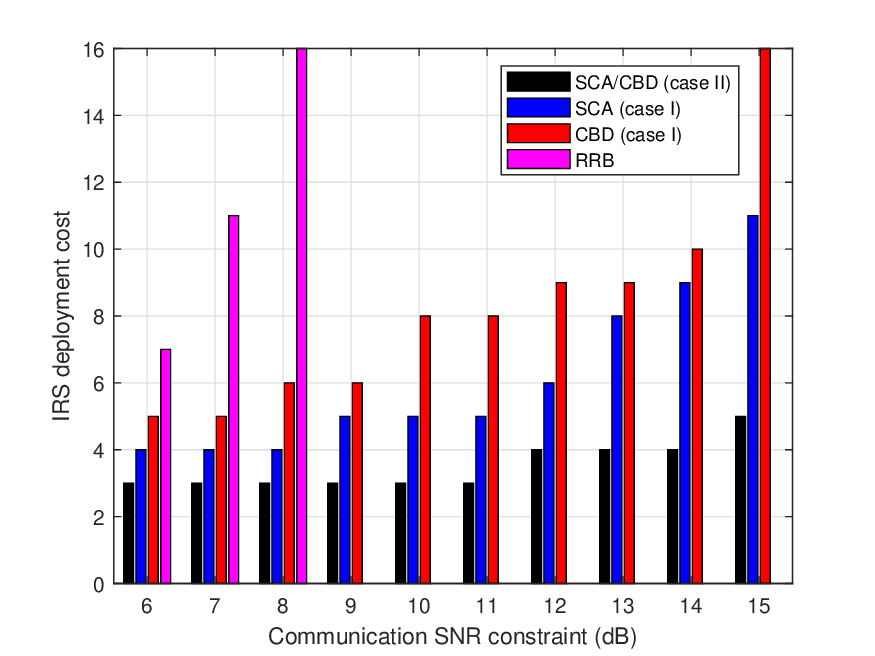}}
	\subfigure[IRS deployment cost versus \(P_\text{s}\) with \(\Gamma_\text{c} = 12\) dB.]
	{\includegraphics[width=0.8\columnwidth]{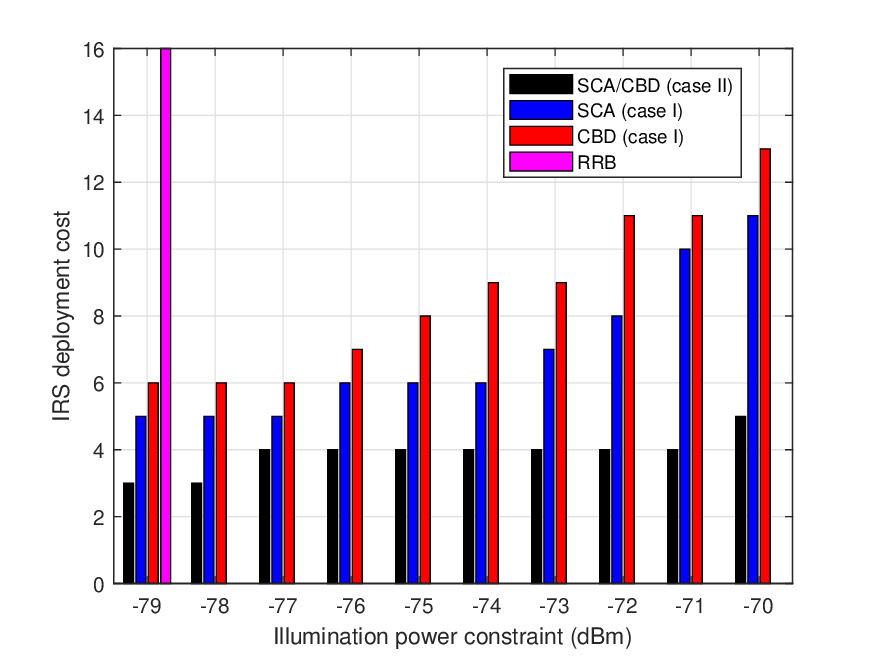}}
	\caption{IRS deployment cost versus communication SNR and sensing illumination power constraints with \(w_2 = 0\).}
	\label{fig_Gam}
\end{figure}
Fig. \ref{fig_Gam} shows the IRS deployment cost under different communication SNR and sensing illumination power constraint \(\Gamma_\text{c}\) and \(P_\text{s}\), respectively. Here, we set the BS power consumption weight as \(w_2 = 0\), i.e., the system tends to deploy as less IRSs as possible, regardless of the power consumption.
It is observed that in case I, our proposed SCA design outperforms the CBD design. Considering that they share the same rounding process, such a performance gap comes from the procedure for determining IRS deployment weight.
It is also observed that in case II, both SCA and CBD designs achieve the same IRS deployment cost. In this case, the location of each IRS is not dominant to its coverage performance due to the flexibility of its dynamic reflective beamforming, and directly increasing the number of deployed IRSs can satisfy higher \(\Gamma_\text{c}\) and \(P_\text{s}\).
Furthermore, compared to SCA and CBD, the RRB design is observed to achieve a much higher IRS deployment cost, even though it optimally determines the IRS deployment. This demonstrates the importance of IRS reflective beamforming design in reducing the system cost.


\begin{figure}[tb]
	\centering {\includegraphics[width=0.4\textwidth]{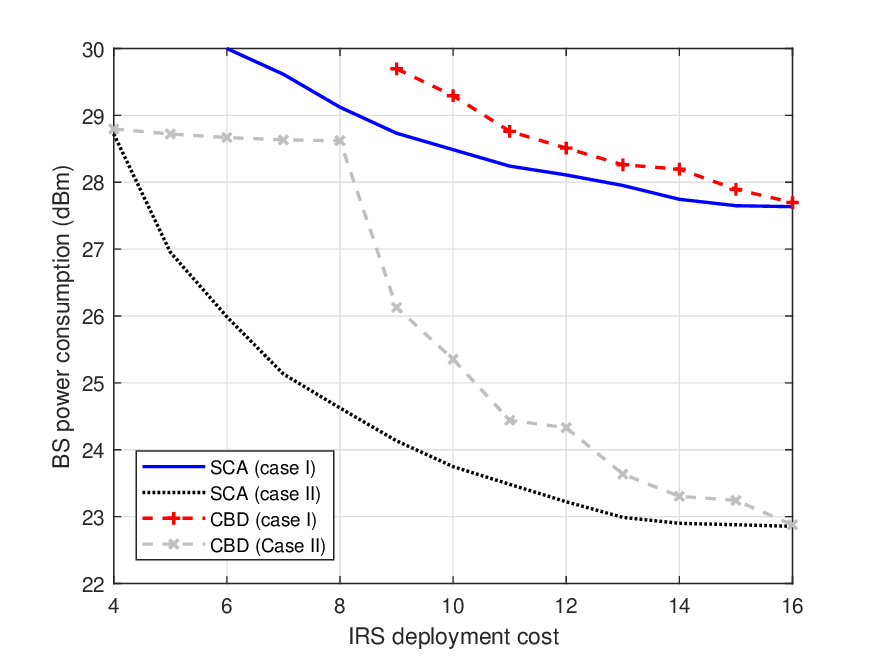}} 
	\caption{BS power consumption versus IRS deployment cost with \(P_\text{s} = -74\) dBm and \(\Gamma_\text{c} = 12\) dB.}
	\label{fig_weight}
\end{figure}
Fig. \ref{fig_weight} shows the IRS deployment cost versus the BS power consumption cost under \(P_\text{s} = -74\) dBm and \(\Gamma_\text{c} = 12\) dB. Such a tradeoff is characterized by changing the value of \(w_2\). 
It is observed that in both cases I and II, our proposed SCA design outperforms the CBD design. This is because with optimized IRS deployment weights, the rounding process of SCA results in a more reasonable IRS deployment indicator, thereby reducing the BS power consumption with the same number of deployed IRSs. Consequently, such performance gaps diminish when the IRSs are fully deployed.
Besides, under this setup, the RRB benchmark is infeasible, which is consistent with the results in Fig. \ref{fig_Gam}.

\begin{figure}[tb]
	\centering 
	\subfigure[IRSs deployment (\(P_0 = 29.63\) dBm).]
	{\includegraphics[width=0.8\columnwidth]{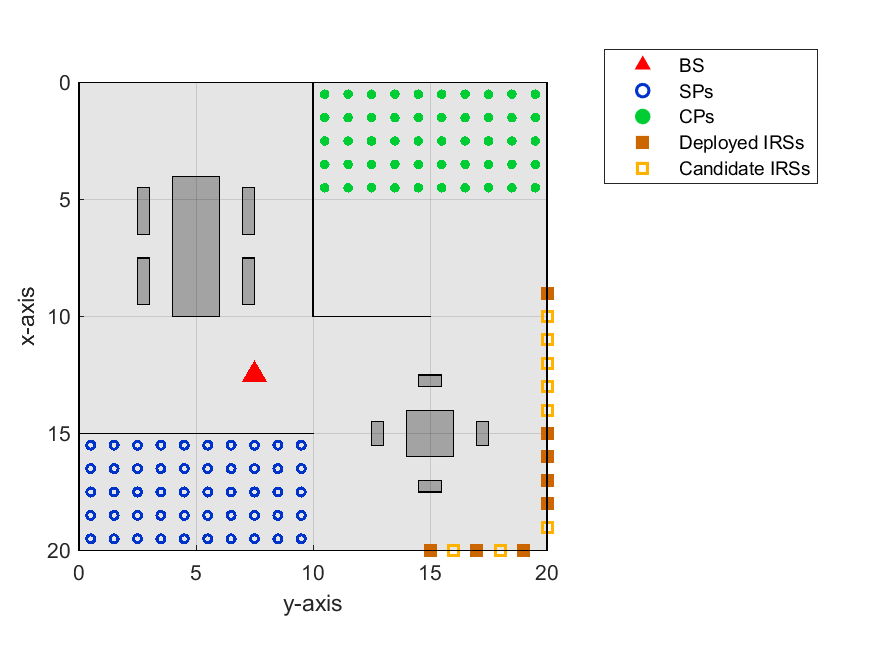}}
	\subfigure[Coverage over the SPs and CPs.]
	{\includegraphics[width=0.9\columnwidth]{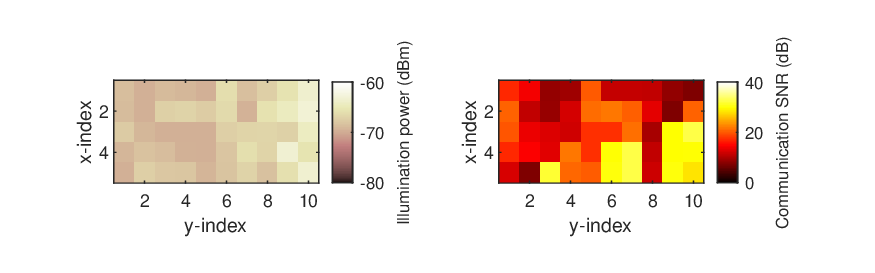}}
	\caption{IRS deployment and coverage performance in case I with \(P_\text{s} = -70\) dBm, \(\Gamma_\text{c} = 8\) dB, and \(w_2 = 1\).}
	\label{fig_hll}
\end{figure}

Finally, to provide more insights, we demonstrate the achieved IRS deployment and the corresponding sensing and communication coverage performance by our proposed SCA design.
Fig. \ref{fig_hll} shows the IRS deployment and coverage performance in case I, with \(P_\text{s} = -70\) dBm, \(\Gamma_\text{c} = 8\) dB, and weight \(w_2 = 1\). 
In this case, the sensing requirement dominates the communication one, and the IRS deployment cost dominates the BS transmit power.
Consequently, to reduce the IRS deployment cost, the BS transmit power almost reaches the maximum power budget. It is observed in Fig. \ref{fig_hll}(a) that the system tends to deploy more IRSs near the sensing area to create more LoS links, while merely one IRS is deployed near the communication area. In addition, with the assistance of deployed IRSs, it is observed in Fig. \ref{fig_hll}(b) that compared to Fig. \ref{fig_0}(b), the illumination power at all SPs and the communication SNR at blocked CPs are markedly enhanced.
  
\begin{figure}[tb]
 	\centering 
 	\subfigure[IRSs deployment (\(P_0 = 29.79\) dBm).]
 	{\includegraphics[width=0.8\columnwidth]{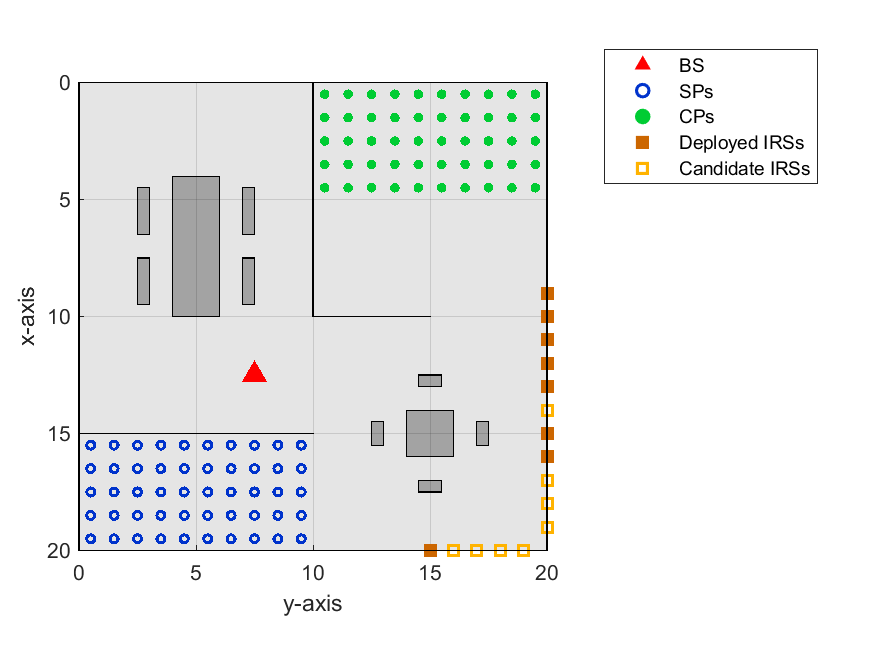}}
 	\subfigure[Coverage over the SPs and CPs.]
 	{\includegraphics[width=0.9\columnwidth]{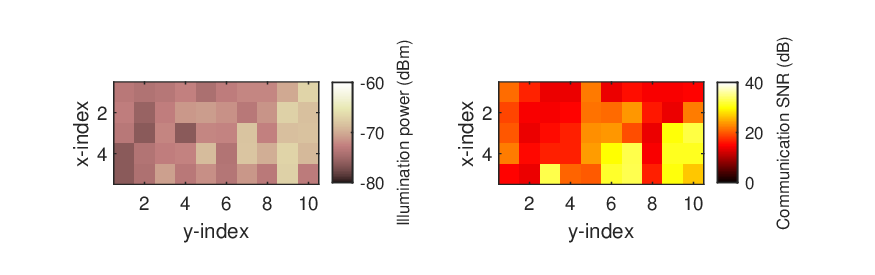}}
 	\caption{IRS deployment and coverage performance in case I with \(P_\text{s} = -76\) dBm, \(\Gamma_\text{c} = 14\) dB, and \(w_2 = 1\).}
 	\label{fig_lhl}
\end{figure}
Fig. \ref{fig_lhl} shows the IRS deployment and coverage performance in case I, with \(P_\text{s} = -76\) dBm, \(\Gamma_\text{c} = 14\) dB, and weight \(w_2 = 1\).
Compared to the sensing requirement, the communication requirement is dominant in this case. Accordingly, it is observed in Fig. \ref{fig_lhl}(a) that compared to Fig. \ref{fig_hll}(a), more IRSs are deployed near the communication area, while a few IRSs are sufficient to guarantee the sensing coverage. In addition, it is observed in Fig. \ref{fig_lhl}(b) that compared to Fig. \ref{fig_hll}(b), the coverage across the SPs and CPs changes along with the IRS deployment. 

\begin{figure}[tb]
	\centering 
	\subfigure[IRSs deployment (\(P_0 = 27.51\) dBm).]
	{\includegraphics[width=0.8\columnwidth]{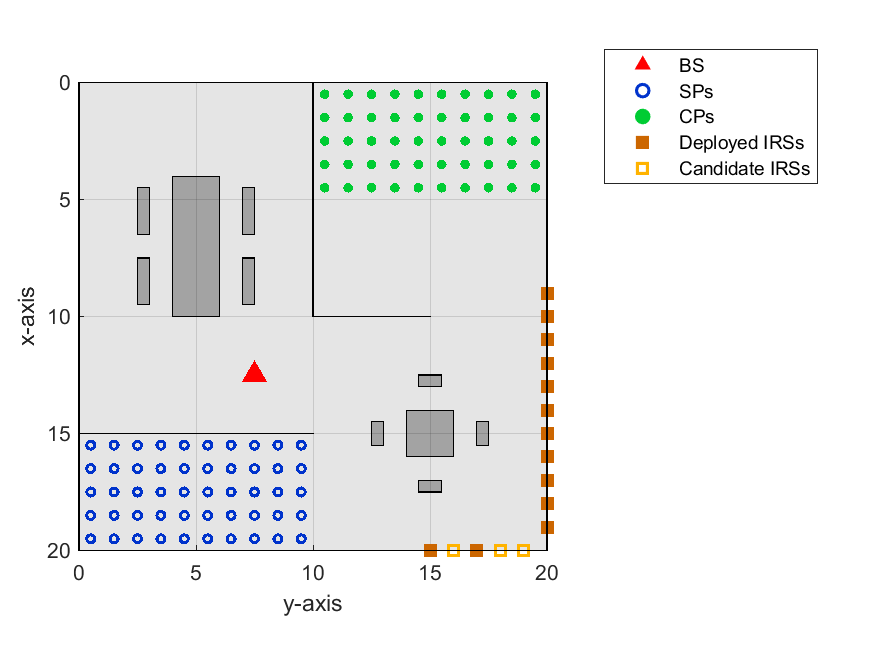}}
	\subfigure[Coverage over the SPs and CPs.]
	{\includegraphics[width=0.9\columnwidth]{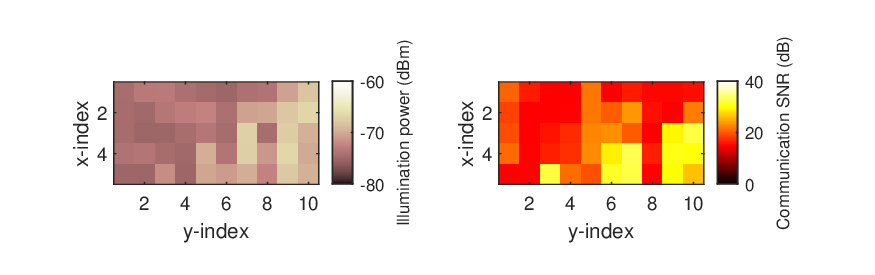}}
	\caption{IRS deployment and coverage performance in case I with \(P_\text{s} = -76\) dBm, \(\Gamma_\text{c} = 14\) dB, and \(w_2 = 100\).}
	\label{fig_lhh}
\end{figure}
Fig. \ref{fig_lhh} shows the IRS deployment and coverage performance in case I, with \(P_\text{s} = -76\) dBm, \(\Gamma_\text{c} = 14\) dB, and weight \(w_2 = 100\). In this case, as \(w_2\) increases, the BS power consumption becomes dominant in the total system cost. Consequently, compared to Fig. \ref{fig_lhl}(a), it is observed in Fig. \ref{fig_lhh}(a) that the system tends to deploy more IRSs, and the BS transmit power is reduced by \(2.28\) dB (\(40.8 \%\)). Moreover, Fig. \ref{fig_lhh}(b) shows the achieved sensing and communication coverage. It is observed that with reduced transmit power, an increased number of deployed IRSs achieves a similar coverage over the SPs and CPs as that in Fig. \ref{fig_lhl}(b).

\begin{figure}[tb]
	\centering 
	\subfigure[IRSs deployment (\(P_0 = 29.84\) dBm).]
	{\includegraphics[width=0.8\columnwidth]{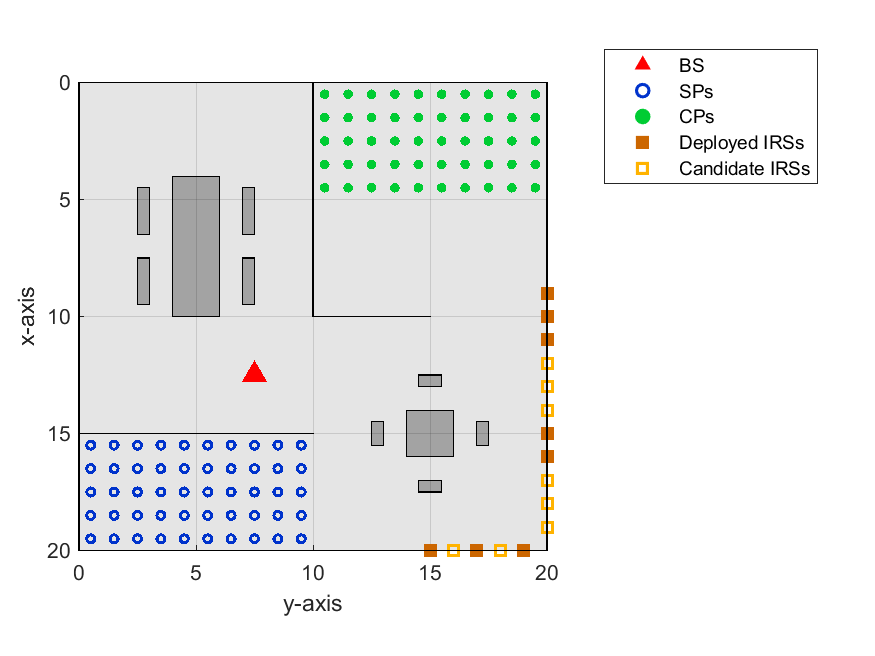}}
	\subfigure[Coverage over the SPs and CPs.]
	{\includegraphics[width=0.9\columnwidth]{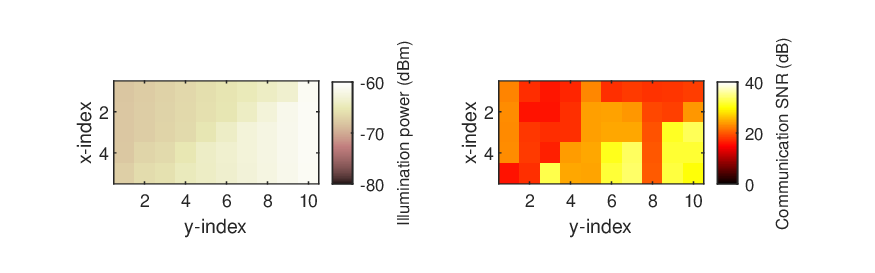}}
	\caption{IRS deployment and coverage performance in case II with \(P_\text{s} = -68\) dBm, \(\Gamma_\text{c} = 16\) dB, and \(w_2 = 1\).}
	\label{fig_hhl}
\end{figure}
Fig. \ref{fig_hhl} shows IRS deployment and coverage performance in case II, with \(P_\text{s} = -68\) dBm, \(\Gamma_\text{c} = 16\) dB, and weight \(w_2 = 1\). In this case, compared to Figs. \ref{fig_hll} and \ref{fig_lhl}, both sensing and communication requirements are significantly increased. However, with the ability to dynamically adjust the reflective beamforming along with the BS beamforming toward each SP and CP, the total number of deployed IRSs does not increase. This validates the performance gap arising from physical constraints compared to the ideal scenario, thereby underscoring the importance of incorporating real-time IRS reflective beamforming designs in practical implementations.


\section{Conclusions}

In this paper, we studied the joint IRS deployment and transmit/reflective beamforming optimization to guarantee the sensing and communication coverage performances while minimizing the system cost in complex propagation environments by leveraging the CKM. To reveal the effect of practical IRS reconfiguration limits on the system performance, we considered two cases with dynamic and quasi-stationary IRS operations. For both cases, we proposed effective solutions based on the relax-and-round method.
Numerical results demonstrated that our proposed SCA-based design properly balances the capital and operational costs by adaptively deploying candidate IRSs according to various sensing and communication coverage requirements.
As a result, it significantly outperforms the CBD heuristic design and the RRB benchmark, highlighting its effectiveness in determining the deployment of candidate IRSs and transmit/reflective beamforming for achieving a superior performance-cost tradeoff.
Besides, with dynamic reflective beamforming, the IRSs achieve significantly enhanced sensing and communication coverage compared to the quasi-stationary case, which underscores the importance of real-time IRS reconfiguration design in practical IRS deployment implementations.

\bibliographystyle{IEEEtran}
\bibliography{ref_IRS}

\end{document}